\documentclass[a4paper,fleqn]{cas-sc}
\usepackage[export]{adjustbox}
\usepackage{amsmath}
\usepackage{etoolbox} 
\usepackage{algpseudocode}

\usepackage{booktabs}
\usepackage{multirow}
\usepackage{siunitx}
\usepackage{lipsum,booktabs}
\usepackage{amssymb}
\usepackage{pifont}
\newcommand{\cmark}{\ding{51}}%
\newcommand{\xmark}{\ding{55}}%
\usepackage{amsmath,amsfonts}
\usepackage{algorithm}
\usepackage{array}
\usepackage[caption=false,font=normalsize,labelfont=sf,textfont=sf]{subfig}
\usepackage{textcomp}
\usepackage{stfloats}
\usepackage{url}
\usepackage{verbatim}
\usepackage{graphicx}
\usepackage{orcidlink}
\usepackage[authoryear,longnamesfirst]{natbib}

\def\tsc#1{\csdef{#1}{\textsc{\lowercase{#1}}\xspace}}
\tsc{WGM}
\tsc{QE}
\tsc{EP}
\tsc{PMS}
\tsc{BEC}
\tsc{DE}

\usepackage{soul}
\usepackage{color}

\begin{document}
\let\WriteBookmarks\relax
\def\floatpagepagefraction{1}
\def\textpagefraction{.001}

\shorttitle{Decentralized and Robust Privacy-Preserving Model}

\shortauthors{Reza Fotohi et~al.}

\title [mode = title]{Decentralized and Robust Privacy-Preserving Model Using Blockchain-Enabled Federated Deep Learning in Intelligent Enterprises}                      



%
\author[1]{Reza Fotohi}[type=,
                        auid=,bioid=,
                        prefix=,
                        role=,
                        orcid=0000-0002-1848-0220]
\fnmark[1]

\ead{r_fotohi@sbu.ac.ir}


\credit{Writing – original draft, Data curation, Writing – review \& editing, Investigation, Conceptualization, Visualization, Software, Validation, Methodology}

\affiliation[1]{organization={Faculty of Computer Science and Engineering},
    addressline={Shahid Beheshti University}, 
    city={Tehran},
    postcode={1983969411}, 
    country={Iran}}

\author[2]{Fereidoon Shams Aliee}[type=,
                        auid=,bioid=,
                        prefix=,
                        role=,
                        orcid=0000-0002-9038-1577]
\cormark[1]
\fnmark[2]

\ead{f_shams@sbu.ac.ir}


\credit{Writing – review \& editing, Supervision, Conceptualization, Formal analysis, Software}
\affiliation[2]{organization={Faculty of Computer Science and Engineering},
    addressline={Shahid Beheshti University}, 
    city={Tehran},
    postcode={1983969411}, 
    country={Iran}}

\author[3]{Bahar Farahani}[type=,
                        auid=,bioid=,
                        prefix=,
                        role=,
                        orcid=0000-0002-7016-6853]
\fnmark[3]
\ead{b_farahani@sbu.ac.ir}

\credit{Writing – review \& editing, Supervision, Conceptualization, Formal analysis, Data curation}

\affiliation[3]{organization={Cyberspace Research Institute},
    addressline={Shahid Beheshti University}, 
    city={Tehran},
    postcode={1983969411}, 
    country={Iran}}

\cortext[cor1]{Corresponding author}



\begin{abstract}
In Federated Deep Learning (FDL), multiple local enterprises are allowed to train a model jointly. Then, they submit their local updates to the central server, and the server aggregates the updates to create a global model. However, trained models usually perform worse than centralized models, especially when the training data distribution is non-independent and identically distributed (non-IID). Because non-IID data harms the accuracy and performance of the model. Second, due to the centrality of federated learning (FL) and the untrustworthiness of enterprises, traditional \emph{FL} solutions are vulnerable to security and privacy attacks. Therefore, to tackle this issue, we propose $\textsc{FedAnil}$, a secure blockchain-enabled \underline{Fed}erated Deep Le\underline{A}r\underline{ni}ng Mode\underline{l} that improves enterprise models' decentralized, performance, and tamper-proof properties, including two main phases. The first phase is proposed to address the non-IID challenge (label and feature distribution skew). In this phase, local models with similar data distributions are grouped into homogeneous clusters using the cosine similarity (CS) and affinity propagation (AP) techniques. Then, for each homogeneous cluster, Wasserstein Generative Adversarial Networks (WGAN) are used to deal with label and feature distribution skew. The second phase was adopted to address security and privacy concerns against poisoning and inference attacks via three steps. In the first step, data poisoning attacks are prevented by using CS. Then, in the second step, collude attacks were prevented by randomly selecting enterprises in the consortium blockchain. Finally, in the third step, model poisoning, membership inference, and reconstruction attacks have been prevented using the CKKS Fully Homomorphic Encryption (CKKS-FHE) technique and consortium blockchain. Extensive experiments were conducted using the Sent140, Fashion-MNIST, FEMNIST, and CIFAR-10 new real-world datasets to evaluate $\textsc{FedAnil's}$ robustness and performance. The simulation results demonstrate that $\textsc{FedAnil}$ satisfies FDL privacy-preserving requirements. In terms of convergence analysis, the model parameter obtained with the $\textsc{FedAnil}$ converges to the optimum of the model parameter. In addition, it performs better in terms of accuracy (more than 11, 15, and 24\%) and computation overhead (less than 8, 10, and 15\%) compared with baseline approaches, namely $\textsc{ShieldFL}$, $\textsc{RVPFL}$, and RFA, respectively. The $\textsc{FedAnil}$ source code can be found on GitHub\footnote{Code available on GitHub Repository: https://github.com/rezafotohi/FedAnil. For any questions about the code, please contact Fotohi.reza@gmail.com}.
\end{abstract}



\begin{keywords}
Privacy-preserving\sep Federated Learning\sep Poisoning Attacks\sep Inference Attacks\sep Intelligent Enterprises
\end{keywords}

\maketitle

\section{Introduction}
\label{ID1}
Data plays a vital role in all decision-making processes and adopting strategies in the current era. Due to the daily increase in the volume of data in intelligent enterprises, analyzing and gaining data insight has become very important. However, operations to obtain data insights are often extracted by models based on machine learning (ML). Because the traditional centralized training of ML models faces security and privacy challenges, a leading technique called Federated Deep Learning (FDL) \citep{ref303} has been developed to solve these challenges \citep{ref22}. FDL is a leading technique based on secure distributed ML that trains a joint deep learning model on heterogeneous clients' data in collaboration with multiple local clients. In this technique, all clients' private data remains confidential, and local clients' updated weights are sent to the central server \citep{ref20}. In FDL, minimizing the sum of the loss function in \emph{n} enterprises is as \eqref{eq7}:

\begin{equation}
\begin{split}
\underset {\boldsymbol\omega\in \mathbb{R}^d }{\min}\hspace{0.1cm} \Biggl\{F(\boldsymbol\omega)=\frac{1}{n} \ast \sum _{k=1}^{n}F^k(\boldsymbol\omega)\Biggl\}.
\label{eq7}
\end{split}
\end{equation}

In \eqref{eq7}, $F^k(\boldsymbol\omega)$ is defined as the enterprise \emph{k} loss/cost function and a function of the dataset $DS_k=\bigl\{X_k, Y_k \bigl\}_{k=1}^{n_k}$ which is held by it. $F^k(\cdot)$ can be stochastic with $F^k(\boldsymbol\omega):=\mathbb{E}_{\zeta_k} F\bigl(\boldsymbol\omega,\zeta_k\bigl)$, where $\zeta_k \in DS_k$ is the sample data in enterprise \emph{k}. To make the concept simpler, we assume that $|DS_{k=1}|=|DS_{k=2}|$ for $k=1$ and $k=2$. This condition does mean that the data distribution of all enterprises is non-IID. In the \emph r$^{th}$ round (with time step $\delta$), the server's standard distributed gradient update will be as \eqref{eq8}:

\begin{equation}
\begin{split}
\boldsymbol\omega_{{r+1}}^{S}=\Bigl(\boldsymbol\omega_{{r}}^{S}- \frac{\delta}{n}\sum _{k=1}^{n}\nabla{F^k\Bigl(\boldsymbol\omega_{{r}}^{S}\Bigl)}\Biggl),
\label{eq8}
\end{split}
\end{equation}

where $\nabla{F^k(\boldsymbol\omega_{{r}}^{S})}$ is calculated using $DS_{k}$ in enterprise \emph{k}, therefore, for the model accuracy to reach the desired threshold, a large number of iterations are executed. However, fewer data are transferred from enterprises to the server. One of the famous lightweight algorithms that have emerged to deal with the challenge of FDL is the federated averaging ({\textsc{FedAvg}}) algorithm. In {\textsc{FedAvg}}, by performing a local training phase, the problem of data distribution among enterprises is solved (according to \eqref{eq9}):

\begin{equation}
\begin{split}
\boldsymbol\omega_{_{h+1}}^{k}=\Biggl(\boldsymbol\omega_{_{h}}^{k}- \delta \nabla{F^k(\boldsymbol\omega_{_{h}}^{k})}\Biggl),\hspace{0.2cm}&{\text{for}}\ \hspace{0.1cm}1 \leq h \leq LT_{r}-1.
\label{eq9}
\end{split}
\end{equation}

In \eqref{eq9}:
\vspace{-\topsep}
\begin{itemize}
  \setlength{\parskip}{0pt}
  \setlength{\itemsep}{0pt plus 1pt}
\item $\boldsymbol\omega^k$: The enterprise \emph{k} model. 
\item $LT_{r}$: Local training time in round \emph{r}.
\item Assuming that $\boldsymbol\omega_{_{0}}^{k}=\boldsymbol\omega_{{r-1}}^{S}$ during round \emph{r}, where the server/global model is represented by $\boldsymbol\omega^{S}$.
\end{itemize}
\vspace{-\topsep}

According to \eqref{eq10}, aggregation of models is done on the received local models.
\begin{equation}
\begin{split}
\boldsymbol\omega_{{r}}^{S}=\Biggl(\frac{1}{n}*\sum _{k=1}^{n}\boldsymbol\omega _{_{_{LT{_r}}}}^{k}\Biggl).
\label{eq10}
\end{split}
\end{equation}

Next, the server transmits the updated global model to the enterprises. As intuition, assuming $LT_{r}=\gamma$ for all rounds of \emph{r}, {\textsc{FedAvg}} on the enterprises side will be \eqref{eq11}:
\begin{equation}
\begin{split}
{\boldsymbol\omega_{r}^{k}} =\begin{cases}
\frac{1}{n}\sum _{_{k=1}}^{n}\boldsymbol\omega_{{r-1}}^{k},&{\text{if}}\ \emph r\hspace{0.1cm}mod\hspace{0.1cm}\gamma=0\\ \\
{\boldsymbol\omega_{{r-1}}^{k}-\delta\nabla{F^k\Bigl(\boldsymbol\omega_{{r-1}}^{k}\Bigl)},}&{\text{otherwise.}} 
\end{cases}
\label{eq11}
\end{split}
\end{equation}

If the averaged vector is given with $\overline{\boldsymbol\omega}_{{r}}:=\frac{1}{n}\sum _{_k=1}^{n}\boldsymbol\omega_{{r}}^{k}$, the number of iterations for the average model will be \eqref{eq12}:
\begin{equation}
\begin{split}
\overline{\boldsymbol\omega}_{{r}}=\overline{\boldsymbol\omega}_{{r-1}}-\biggl(\frac{\delta}{n}\sum _{k=1}^{n}\nabla{F^k\Bigl(\boldsymbol\omega_{{r-1}}^{k}\Bigl)\biggl)},\hspace{0.3cm}&{\text{if}}\ \emph r\hspace{0.1cm}mod\hspace{0.1cm}\gamma\neq0.
\label{eq12}
\end{split}
\end{equation}

Assume that $\parallel$ $\nabla{F^k(\boldsymbol\omega)}$$\parallel$$^2$$\leq C$. The divergence of the $\boldsymbol\omega^{k}$ model of each enterprise from the $\overline{\boldsymbol\omega}_{{r}}$ average model will be according to \eqref{eq13}:
\begin{equation}
\begin{split}
\Vert \boldsymbol\omega_{{r}}^{k}-\overline{\boldsymbol\omega}_{{r}}\Vert^2 \hspace{0.1cm}\leq \hspace{0.1cm}\delta^2\hspace{0.04cm}\overline{r}^2\hspace{0.04cm}C.
\label{eq13}
\end{split}
\end{equation}

Therefore, decreasing the variable $\overline{r}$ makes the {\textsc{FedAvg}} algorithm behave like a standard {\textsc{Gradient Descent}}. On the contrary, if the variable $\overline{r}$ increases, the convergence probability of {\textsc{FedAvg}}  will decrease, and the communication overhead will increase.

However, researchers have demonstrated concerns about non-IID data, security, and privacy in the \emph{FL} process.

The challenge of non-IID data among clients leads to divergence and increases the bias of local models, decreasing the global model's accuracy. Therefore, various approaches have been considered in the research to solve the non-IID data challenge: Clustering, Calibrating the logits, etc. Gu et al. \citep{ref330} proposed a clustering algorithm called FedRC, which is used to overcome feature and label skews. The proposed clustering combines a two-level optimization problem and an objective function. In another work by Wang et al. \citep{ref340}, they proposed a method to solve the non-IID data, especially the label distribution skew, which is called FedBalance. FedBalance focuses on the label skew, a common scenario in data heterogeneity where the label classification of data in each client is unbalanced. Optimization bias is corrected among local models by calibrating the logits in FedBalance.

Due to security and privacy concerns, the local enterprise's data may be poisoned, producing a poisoned model. Furthermore, a central server may infer or reconstruct the local enterprise's data through the local model gradients. Therefore, various approaches have been considered to solve security and privacy concerns. Yazdinejad et al. \citep{ref201} proposed an approach called AP2FL, which is applied in healthcare. To reduce data privacy concerns, they proposed a strategy based on trusted execution environments (TEE) in which model training and aggregation are done thoroughly and securely. In the TEE-based AP2FL model, there is no concern about model aggregation because intruders cannot access gradients. This approach resists privacy attacks such as data reconstruction and parameter inference attacks. Wang et al. \citep{ref301} introduced a comprehensive blockchain-based defense mechanism that is resistant to gradient inversion and poisoning attacks. This mechanism is a combination of public and private blockchains that work together to prevent attackers from reconstructing images using obtained gradients. This preserves the privacy of local model gradients. To solve the privacy problems of federated learning, researchers have proposed different approaches to make the parameters of each local enterprise invisible to the central server during the training process. However, these approaches pose challenges against poisoning attacks \citep{ref304}. In poisoning attacks, a malicious local enterprise sends malicious parameters to update the global model during the training process \citep{ref305}\citep{ref306}\citep{ref307}. These attacks can undermine the performance (accuracy) of federated learning models. In federated learning-based approaches, the central server cannot distinguish between benign and poisoned gradients due to the invisibility of data features from participants, which ultimately leads to training failure. Therefore, preserving the privacy of local enterprise models and defending against poisoning attacks have become critical issues that need to be addressed. A secure aggregation algorithm called Krum is proposed to avoid poisoning attacks by Blanchard et al. \citep{ref302}. In Krum, malicious gradients must be far from Benign gradients to poison the global model. Krum selects the $n-m-2$ closest neighboring gradients in Square Euclidean norm space from a set of input gradients. $n$ is the total number of clients, and $m$ is the upper bound on the number of malicious clients in the federated learning system.

In general, three techniques have been proposed to solve privacy problems in \emph{FL}: Homomorphic encryption (HE), Differential Privacy (DP), and Secure Multiparty computation (SMC). An approach based on HE to guarantee the confidentiality of gradients is proposed by Bonawitz et al. \citep{ref310}. In this approach, the server updates the parameters and sends them to local participants to decrypt and update the model. Aono et al. \citep{ref309} used HE to preserve the privacy of local models so that local models can be exchanged between servers and clients without decryption. Similarly, a privacy-preserving approach based on HE is proposed by Zhang et al. \citep{ref311}, which encrypts updates by \emph{HE} and uses the DSSGD method to enable distributed encryption in the local training phase. To secure aggregation, a HE-based approach is proposed by Li et al. \citep{ref312}, which aims to make \emph{FL} robust against inference attacks. Unfortunately, if malicious participants fake a large false value, it will not be efficient. In summary, the above schemes aim to make individual gradients invisible to the central server to protect privacy. In DP-based approaches, inference attacks are prevented by adding noise to the data. Still, this technique is not resistant to data reconstruction attacks \citep{ref308}. In a similar work by Zhao et al., \citep{ref313}, they used differential privacy to address the privacy problem and defend against poisoning attacks. However, the use of differential privacy reduces the accuracy of the model. The third category is related to the MPC technique. Li et al. \citep{ref304} proposed an MPC-based approach to defending against poisoning attacks, which requires multiple communications in a multi-party multiplication operation. A similar work by Liu et al. \citep{ref314} used the correlation coefficient to evaluate the training quality of each participant but required a trusted third party to generate secure parameters. Furthermore, the above schemes mainly defend against poisoning attacks by exclusively removing malicious gradients without the ability to identify malicious participants.

However, the FDL model trained in local enterprises must have high accuracy on non-IID data. Second, it should be robust to poisoning and inference attacks. Therefore, this paper faces the following two major challenges:
\begin{enumerate}
\item \emph{non-IID}: Since the training data in enterprises are collected by themselves based on the usage pattern and local environment, the \emph {feature and label} skews usually differ among enterprises. Local enterprise models try to get closer to the local optimum, which may move them away from the global optimum. According to Fig. \ref{fig_Unbalanced_and_non-IID}, in the non-IID setting (Fig.\ref{non-IID}), the average local model ($\overline{\boldsymbol\omega}_{_{r}}$) divergence is higher than the global model (${\boldsymbol\omega}_{r}^{S}$). But in the IID setting (Fig.\ref{IID}), the average local model ($\overline{\boldsymbol\omega}_{_{r}}$) is closer to the global model (${\boldsymbol\omega}_{r}^{S}$). Note that the more communication rounds there are, the greater the divergence in the non-IID setting. Therefore, non-IID settings can negatively affect the accuracy and convergence of a global model \citep{ref14,ref15}.

\begin{figure*}[!t]
\centering
\subfloat[]{\includegraphics[width=2.0in]{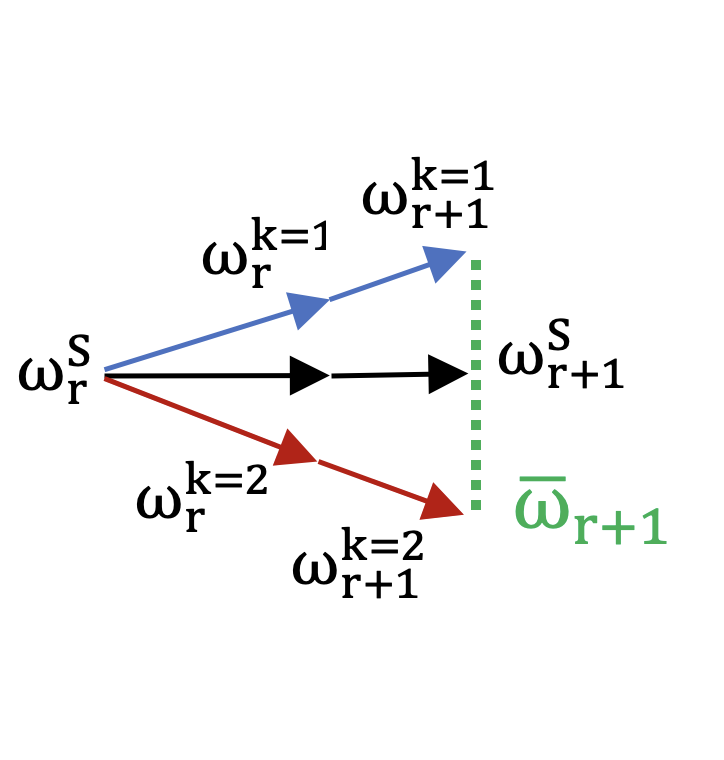}%
\label{IID}}
\hfil
\subfloat[]{\includegraphics[width=2.0in]{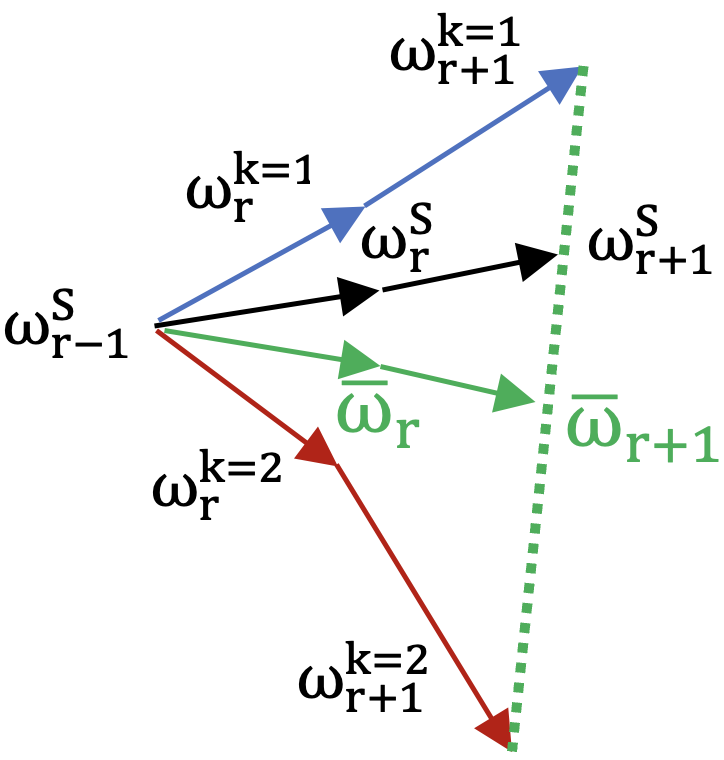}%
\label{non-IID}}
\caption{Overview of drift in FDL under IID and non-IID settings, where $\boldsymbol\omega_{r}^{S}$ is the global model and $\overline{\boldsymbol\omega}_{_{r}}$ is the averaged model of the local enterprise $1$ ($\boldsymbol\omega_{r}^{k=1}$) and local enterprise $2$ ($\boldsymbol\omega_{r}^{k=2}$). (a) IID. (b) non-IID.}
\label{fig_Unbalanced_and_non-IID}
\end{figure*}

\item \emph{Privacy-preserving}: FDL-based techniques emerged to solve the privacy challenge. This technique sends no raw data from the local enterprises to the server. However, FDL does not always guarantee sufficient privacy, as inference attacks violate it. In an inference attack, intruders infer the training data adopted to train the ML-based model through the model parameters to violate privacy. On the other hand, data and model poisoning attacks are other concerns in FDL that aim to reduce global model accuracy. Therefore, in data poisoning, the attacker manipulates the training dataset. Furthermore, manipulating the gradient value the optimizer calculates in the model poisoning attack de-optimizes the model to increase the test error rate \citep{ref15, ref20}.
\end{enumerate}

Therefore, the main focus of this paper is \emph{how the proposed model can reduce the security and privacy concerns of training data and local enterprise's model parameters against poisoning and inference attacks.} To address these concerns, we propose $\textsc{FedAnil}$, a privacy-preserving federated deep learning model that enables servers to aggregate local models securely. In addition, by using the \emph{CS} technique, suspicious models are identified and excluded from participating in the global model aggregation process. Our main contributions are unfolded as follows:
\begin{enumerate}
\item \emph{non-IID}: In the $\textsc{FedAnil}$ model, on the server side, local models with similar data distributions are grouped into homogeneous clusters using \emph{CS} and \emph{AP} techniques. Then, Wasserstein Generative Adversarial Networks (WGAN) were used for each homogeneous cluster to tackle the label and feature distribution skew. In this step, a generator and a detector (global model) try to learn $\vartheta$. Therefore, the global model by training in an adversarial process can detect $\vartheta$ that causes a decrease in accuracy and convergence in the FDL-based model {\bf\emph{(discussed in Section \ref{FM_Unbalanced_non-IID})}}.
\item \emph{Privacy-preserving}: In the first step, data poisoning attacks are prevented by using \emph{CS}. In this technique, the updated gradients of each local enterprise are measured in terms of the normality of their updated gradients compared to the prior round global model, and any gradients with abnormal values are removed. Suppose this operation is repeated \emph{n} times in \emph{r} different rounds; the enterprise will be considered an intruder, and its data will be considered poisoned and removed from participation in the FDL process. Then, in the second step in FDL, collude attacks have been prevented by randomly selecting enterprises in the consortium blockchain. Finally, model poisoning attacks, membership inference, and reconstruction can be prevented in the third step using the Cheon-Kim-Kim-Song (CKKS-FHE) \citep{ref31} technique and consortium blockchain. In this step, on the local enterprise's side, \emph{CKKS-FHE} is used to encrypt both \emph{CHs index vector} and \emph{initial gradients using the CH index}. On the server side, aggregation operations are performed without model parameter decryption {\bf\emph{(discussed in Section \ref{FM_Privacy-preserving})}}.
\end{enumerate}

The rest of this paper's organization will be as follows: Section \ref{PR1} presents the relevant preliminaries used in the $\textsc{FedAnil}$ model. Next, in Section \ref{RW1}, related work is briefly summarized. We introduce the proposed $\textsc{FedAnil}$ in detail in Section \ref{FM1}. The convergence analysis of $\textsc{FedAnil}$ is discussed in Section \ref{CA1}. Section \ref{EP1} presents the $\textsc{FedAnil}$ experimental result by comparing it with representative models. Finally, in Section \ref{CC1}, conclusions and future work are brought.

\section{Preliminaries}\label{PR1}
Some basic concepts used in $\textsc{FedAnil}$ are reviewed in this section. In this paper, every client can also be called an enterprise. The symbols used are given in Table \ref{tab2} for description clarity.

\subsection{Intelligent Enterprise}
An Intelligent enterprise continuously uses advanced technologies and best practices in integrated and agile business processes. Every intelligent enterprise has the following three main components: {\emph{Intelligent Suite}}, {\emph{Digital Platform}}, and {\emph{Intelligent Technologies}} \citep{ref9}.

\subsection{K-Medoids Clustering}
In this algorithm, which works based on repetition, all data sets are divided into unique subgroups called clusters. Unlike the other famous clustering algorithms, the $\textsc{K-Medoids}$ \citep{ref67} method uses the cluster's most central sample as the cluster's representative instead of getting the average values of the samples. For this reason, this algorithm shows little sensitivity to out-of-range data. The cost in the $\textsc{K-Medoids}$ algorithm is calculated through \eqref{eq1}:

\begin{equation}
 \partial =\sum _{\partial _i} \sum _{P _i \in \partial _i}
 \Big| P_i - \partial _i \Big|.
 \label{eq1}
\end{equation}

In \eqref{eq1}, the dissimilarity of Medoids ($\partial _i$) and objects ($P _i$) is calculated using $\partial$, which is the loss function.

\begin{table}
	\caption{Key notations.}
	\setlength{\tabcolsep}{1.5\tabcolsep}
	\centering
	\begin{tabular}{p{1.3cm}|p{6.3cm}}
		\toprule
		\textbf{Notation} & \textbf{Definition} \\
		\midrule
		$X_{k}$,$Y_{k}$ & Features,\hspace{0.1cm}Label for enterprise \emph{k}\\
            $\boldsymbol\omega^{S}$ & The server/global model\\
		$\boldsymbol\omega(\tau)^k$ & The local model $\tau$ of the enterprise \emph{k}\\
		$\boldsymbol\omega(\tau)^{S}$ & The global model $\tau$ of the server\\
		$LT_{r}$ & Local training time in round \emph{r}\\
		$\Delta c$ & Selected local enterprises for training\\
		\emph{k} & Participating clients; (1 $\leq$
            \emph{k} $\leq$ \emph{n})\\
		\emph{r} & Current communication round\\
  		\emph{R} & Total communication rounds\\
		$\boldsymbol\omega$, $\nabla\boldsymbol\omega$,\emph{n} & Weight, Gradient, Total enterprises\\
		$\varphi_1$, $\varphi_2$ & Thresholds $1$ and $2$ in the \emph{CS}\\
            $\varphi$ & WGAN loss function threshold\\
		$\chi^k$, $\vartheta$ & Malicious enterprise \emph{k}, Hard shadow samples\\
		$\Upsilon$ & Initial gradients using the \emph{CH} index\\
  		$\Psi$ & \emph{CH} and their index\\
		$\tau$,$\mathcal{M}$ & Selected model; ($\tau$ $\in$ $\mathcal{M}$), Models vector\\
		$\iota$ & Total samples in the validation dataset\\
		$\kappa$ & Correctly predicted samples \\
		$\mathbb{G}$, $\mathbb{D}$ & Discriminator, Generator \\  
            $\aleph$ & Local enterprise delay\\
            $\mathbb{GI}$ & Global iteration; ($\aleph + r$)\\
		\bottomrule
	\end{tabular}
	\label{tab2}
\end{table}
\subsection{Privacy-Preserving FL (PPFL)}
PPFL means that in the training process, each client's raw local datasets or local individual model parameters must be securely protected, even if some clients collude \citep{ref22}.

\subsection{FL-based Models Attacks}
Attacking FL systems means that an intruder can cause models to fail to perform specific tasks. Attacks that reduce model accuracy and cause data security and privacy violations are classified into Poisoning and Inference.

Poisoning attacks (training time attacks) aim to breach data security (targeting the integrity dimension of the CIA) and then reduce the global model's efficiency and accuracy. In \emph{FL}, each client has a dataset that may be modified by an attacker for malicious purposes. Since an insider attacker can access the training process, model, and dataset, it can use these capabilities to perform poisoning attacks. When the jointly global model is poisoned, the central server shares it with every local enterprise in the network and poisons all local models \citep{ref68}.

Inference attacks (inference-time attacks) aim to breach data security (targeting the confidentiality dimension of the CIA) and subsequently breach data privacy. Inference attacks use different knowledge to infer private information or sensitive data. Based on the attacker's capabilities and the attack phase, the attacker exploits Oracle's access to the model, its internal computations, or updates\citep{ref68}\citep{ref69}.

Details of the five types of poisoning and inference attacks are explained below:
\vspace{-\topsep}
\begin{enumerate}
  \setlength{\parskip}{0pt}
  \setlength{\itemsep}{0pt plus 1pt}
	\item {\emph{Data poisoning attacks}}. A data poisoning attack aims to manipulate clients' local data to create misclassifications. The attacker modifies the clients' training data sets in this attack to reduce the model's overall performance (\emph{untargeted}). Data poisoning attacks include two general categories: \emph{Noise injection} and \emph{Label-flipping}, which are explained below:

\begin{itemize}
  \setlength{\parskip}{0pt}
  \setlength{\itemsep}{0pt plus 1pt}
  \item {\bf{Noise injection:}} In a noise injection-based data poisoning attack, the attacker can change specific data characteristics by injecting noise. In this attack, the intruder can directly destroy the model's updated parameters by influencing the local FL systems in the data aggregation process. For example, participants have permission to access local training datasets, so attacker participants may add excessive noise that degrades the quality of local datasets and produces unreliable local model updates. This is very easy for attacker participants to do without the knowledge of the central server.
\item {\bf{Label-flipping:}} Another type of data poisoning attack is label-flipping. This attack manipulates the datasets during the local training process to reduce the model's overall performance using the label-flipping technique. The intuition behind the label-flipping technique is to change the labels of a local dataset. Label-flipping attacks can be made in two ways, namely \emph{Clean-Label} and \emph{Dirty-Label}:
\begin{itemize}
  \setlength{\parskip}{0pt}
  \setlength{\itemsep}{0pt plus 1pt}
   \item {\bf{Clean-Label:}}A clean-label attack is a form of data poisoning attack where the attacker only modifies the training data input without needing access to the labeling function. In other words, an attacker cannot change data labels in a dataset, but it puts the poison data samples online and waits for the real client to label them and put them in its dataset. This type of attack occurs when data is collected from untrusted sources and is therefore very difficult to identify Because the poison sample is properly labeled. Hence, the model is trained on a poison dataset and thus becomes infected.
 \item {\bf{Dirty-Label:}}This type involves modifying the dataset by changing the label of a data sample to an optional data sample (targeted) or to a random sample (\emph{untargeted}). For example, changing all airplane labels to birds that the final model of aircraft images classifies as (targeted) birds; therefore, the model is trained on poison data.
  \end{itemize}
  \end{itemize}

{\setlength{\parindent}{15pt}
Therefore, the $\textsc{FedAnil}$ assumes the data poisoning attack is performed through \emph{noise injection}. In the data poisoning attack based on noise injection, the intruder's goal is to inject noise into the training data of local enterprises. Therefore, this attack reduces the accuracy of the trained model to keep the global model from converging}\citep{ref13, ref38, ref15}.However, if this attack is made by label-flipping (\emph{Clean-Label} or \emph{Dirty-Label}), the proposed model will not resist it.

 	\item {\emph{Model poisoning attacks}.}The second attack from the category of poisoning attacks is the model poisoning attack. Generally, two types of model poisoning attacks are based on the attacker's goal: \emph{Targeted} and \emph{Untargeted}. For targeted attacks, the attacker aims to reduce the model's accuracy on some specific samples/inputs. In contrast, \emph{untargeted} attacks aim to minimize the model's accuracy on the data. In targeted attacks, the attacker modifies the behavior of the model only on some selected inputs of the attacker while preserving accuracy in the \emph{FL} task. Targeted attacks cause the model to misclassify a set of specific input samples, while \emph{untargeted} attacks are designed to reduce the accuracy of global model predictions for different samples.
  
  {\setlength{\parindent}{15pt}
  More precisely, \emph{untargeted} attacks are considered model update poisoning attacks. Unlike centralized learning models, the distributed nature of \emph{FL} creates a new threat against poisoning attacks. In model poisoning attacks, the attacker directly manipulates local model gradients. For example, the attacker can inject random noise into the local gradients, change the sign of the updated gradients, or replace the values of the gradients with a fixed value to affect the convergence of the global model. The purpose of model poisoning attacks is to neutralize global model learning. As a result, the global model can be heavily poisoned with invalid poison gradients during the aggregation process. Due to the stronger ability of the attacker, model poisoning attacks can have a greater impact on model accuracy. Therefore, a successful untargeted attack can dramatically reduce the accuracy of the global model.}

{\setlength{\parindent}{15pt}
Therefore, the $\textsc{FedAnil}$ considers the \emph{noise injection-based poisoning model}. If the model's poisoning is done in another way, such as changing the sign of the gradients or replacing the gradients with a constant value, the proposed model will not have a good resistance against it.}

	\item \emph{{Collude attack}.} In this type of poisoning attack, several enterprises deliberately make a secret contract with different enterprises to create an attack against the aggregator server. Collude attacks are divided into two categories, which are explained below \citep{ref38, ref13}:
 \begin{itemize}
  \setlength{\parskip}{0pt}
  \setlength{\itemsep}{0pt plus 1pt}
   \item {\bf{Cross-update Collude:}} The previous round clients can collude with the current round clients to execute attacks on current global model updates.
   \item {\bf{Within-update Collude:}} The current round clients are coordinated to attack the current round global model.
 \end{itemize}

{\setlength{\parindent}{15pt}
In the $\textsc{FedAnil}$, \emph{within-update collude} is considered, and it is assumed that clients do not have the possibility of performing cross-update collude attacks that require coordination with previous rounds.}

	\item \emph{{Membership inference attack}.} The first attack in the category of inference attacks is the membership inference attack, which targets the privacy of local clients. In an inference attack, intruders infer the training data adopted for the ML-based model training through model parameters to violate privacy. Due to black box access, an attacker does not necessarily need knowledge about the internal parameters of the target machine learning model to perform a membership inference attack. An attacker can analyze the classification result of arbitrary data to perform a membership inference attack \citep{ref15}.

{\setlength{\parindent}{15pt}
In the $\textsc{FedAnil}$, a \emph{black box-type membership inference attack} is considered. If the membership inference attack is made in another way, such as a white box, the $\textsc{FedAnil}$ will not have a good resistance against them.}

	\item \emph{{Reconstruction attack}.} The second attack from the category of inference attacks is a reconstruction attack. This attack aims to access sensitive training data inputs by manipulating local gradients received on the server side. In this attack, the attacker tries to reconstruct/extract/infer the sensitive features/inputs of the training data through the trained model (model output) and the insensitive features. In other words, the data reconstruction attack aims to reconstruct the training samples/related labels used in the training process. In this attack, the attacker uses a machine learning model and the incomplete information it has about the data to infer the missing information for that data.
 
 {\setlength{\parindent}{15pt}
 For example, the attacker has detailed information about a person's medical record and uses a trained model to infer the person's Genotype. In particular, the reconstruction attack can be categorized depending on whether the attacker can only search the model, i.e., \emph{black box}, or access model parameters, i.e., \emph{white box}. Specifically, the reconstruction attack attempts to reconstruct a sample of the data $D$ from a model such as $f$ trained on the dataset $D$ by querying $f$}\citep{ref13,ref15}.

{\setlength{\parindent}{15pt}
In the $\textsc{FedAnil}$, a \emph{black box reconstruction attack} is considered. If the reconstruction attack is made in another way, such as a white box, the $\textsc{FedAnil}$ will not have good resistance against them.}

\end{enumerate}
\vspace{-\topsep}

\subsection{Cosine Similarity}
This research uses the \emph{CS} technique to calculate the similarity between gradient vectors. If the similarity between the two vectors is less, the output value is $-1$, and vice versa; if the similarity is higher, the output value is $+1$ \citep{ref11}.

\subsection{Consortium Blockchain}
The difference between the consortium and other types of blockchain is in the consensus mechanism. In the consortium, the consensus is controlled by pre-defined enterprises. To ensure the correct performance, the consortium has a validator node that can perform two actions: Verifying and initiating or receiving transactions \citep{ref12}.

\subsection{Homomorphic Encryption}
This algorithm is an encryption and computation algorithm that performs computation operations on encrypted data. In this algorithm, the output is encrypted. If this algorithm satisfies equation \eqref{eq2}, it will be holomorphic \citep{ref31}:
\begin{equation}
	\begin{split}
		\text E\bigl(text_{1}\bigl)\hspace{0.1cm}*\hspace{0.1cm}E\bigl(text_{2}\bigl) = E\Bigl(text_{1} * text_{2}\Bigl) \hspace{0.2cm}  \forall_{text_{1},text_{2}} \in M.
		\label{eq2}
	\end{split}
\end{equation}

In \eqref{eq2}, the operator $\ast $ denotes a homomorphic operation, and  \emph{M} denotes the set of all possible messages. In this paper, the \emph{CKKS-FHE} is used.

\subsection{L-BFGS optimization algorithm}
The \emph{L-BFGS} algorithm is one of the $\textsc{Quasi-Newton}$ classes, a type of second-order optimization algorithm. \emph{L-BFGS} algorithm is used when direct calculation is impossible, and therefore, it is approximated using the second-order derivative operation. The $\textsc{pseudo-Newton}$ method approximates the inverse Hessian using gradients and can be computationally feasible \citep{ref70}.

In $\textsc{FedAnil}$, the \emph{L-BFGS} algorithm has been used as an optimizer to match the gradients of all training model parameters.

\section{Related Work}\label{RW1}
\emph{FL}-based models are vulnerable to attacks and violate data security and privacy. Also, because the data distribution of enterprises is non-IID, it leads to the divergence of the model, and as a result, the model's accuracy decreases. Due to this issue, many approaches have been developed over the past years to reduce privacy concerns and solve the non-IID challenge (e.g., \citep{ref2}, \citep{ref19}, \citep{ref20}, \citep{ref14}, \citep{ref10}, \citep{ref16}, \citep{ref35}, \citep{ref61}, \citep{ref62}, \citep{ref63}).In this section, we first review some approaches that reduce security and privacy concerns. Then, we will review the most important recent works that addressed the non-IID challenge.

A HE-based Privacy-Preserving approach is proposed in \citep{ref2}. This approach has high resistance against model poisoning attacks and guarantees model privacy. Specifically, this approach uses CS to detect model poisoning attacks, measuring the distance between two encoded gradients. 

In \citep{ref19}, a Privacy-Preserving method is proposed that guarantees the local model's privacy against poisoning attacks. The main idea of this algorithm in identifying the poisoning attack is to measure the direction and magnitude of the gradients in the cipher text mode and remove the suspicious models from the aggregation process. Therefore, the proposed method guarantees the correctness of aggregation of local users' models on the server side, which leads to the Privacy-Preserving of local users' models. 

In \citep{ref20}, a robust federated aggregation-based approach ({\textsc{RFA}}) that is resistant to poisoning attacks is proposed. Two algorithms have been used in {\textsc{RFA}}. The aggregation operation of all local models is performed on the server side using the first algorithm at high speed. The second algorithm performs the personalization of gradients in clients.

One of the most common \emph{FL} averaging methods is {\textsc{FedAvg}}. This algorithm trains a global model by randomly selecting the clients \citep{ref14}. Another algorithm that emerged as an improved {\textsc{FedAvg}} algorithm was the {\textsc{FedProx}}. This algorithm is compatible with non-IID data and has used Euclidean distance to improve the global model efficiency \citep{ref10}. Another popular algorithm, introduced as adaptive server optimization, is {\textsc{FedAdam}}. This algorithm guarantees the model convergence against non-IID data \citep{ref16}.

Several works exist on decentralized FL with and without noisy channels, server/client selection, and dynamic model parameter updates. 

In \citep{ref35}, three decentralized FL algorithms are proposed to solve imperfect communication between different agents. In the first algorithm, noise has been added to the algorithm's parameters to simulate the scenario of noisy communication channels. In addition to adding noise to the algorithm's parameters in the second algorithm, it averages rumors before gradient optimization. Finally, in the third algorithm, noise shares gradients through noisy communication channels instead of parameters. 

Li et al. \citep{ref61} proposed a method called FEDBN for cancer detection without sharing privacy-sensitive data. The main challenge that FEDBN faces is to solve the problem of feature skewness of medical images under a non-IID setting. In FEDBN, the local data regarding distribution in the feature space is skewed, and this scenario is identified as a feature skewness challenge. This type of non-IID data is a critical problem in many real-world scenarios, typically where local clients are responsible for heterogeneity in feature distributions. However, image appearance can vary greatly due to different imaging machines and hospital protocols, such as intensity and contrast. In this method, local batch normalization has been used to reduce feature skewness before averaging the models. In particular, FEDBN keeps the local batch normalization parameters consistent with the global model in the aggregation step to reduce the feature bias in non-IID data.

Chen et al. \citep{ref62} proposed a method called CalFAT to deal with the challenge of label skewness under non-IID data. Label skew leads to unequal class probabilities and heterogeneous local models. In this method, the problem of label skew was studied, and a root cause was revealed, i.e., the instability of local model training and the issues of natural accuracy reduction. In particular, CalFAT is proposed to deal with the instability issue by adaptively calibrating logits to balance classes. Therefore, CalFAT optimization can lead to locally homogeneous models and, as a result, stable training, faster convergence, and better final performance.

Gu et al. \citep{ref63} presented an algorithm called FedBR to solve the feature and label distribution skew. Current approaches are limited by slow and unstable convergence due to data diversity in different clients. In FedBR, this bi-skewness is due to the bias of local updates in FL, where the biased local classifiers cannot classify the unseen data effectively. Therefore, the FedBR algorithm aims to use pseudo-data to reduce the local learning bias of features and classifiers. Specifically, the FedBR algorithm has two main components. The first component helps reduce bias in local classifiers by balancing the output of local models. The second component helps to learn local features similar to global ones, but they differ from those learned from other data sources. This work solves the feature and label distribution skew under the FedBR algorithm.

In \citep{ref64}, a \emph{FL} framework is proposed to solve the non-IID challenge, especially the feature and label distribution skew, called FedFA. A feature and label distribution skew in which different clients have an unbalanced number of features and labels belonging to the target features and labels. The presence of feature and label bias among clients leads to a vicious cycle between classifier divergence and feature inconsistency among client models, which reduces training performance. Specifically, in FedFA, feature anchors are used as a loss function to align features and labels and calibrate classifiers across clients simultaneously. This allows the client models to be updated in the same feature space with fixed classifiers during local training. As a result, the FedFA framework solves the challenge of features and label skews among local models so that the model accuracy does not decrease in the face of these skews, which leads to an increase in efficiency.

In Table \ref{tab1}, all related approaches were compared in terms of robustness to non-IID, privacy-preserving, and security. In summarizing this section, the main differences between the $\textsc{FedAnil}$ model with existing approaches are as follows:

\vspace{-\topsep}
\begin{itemize}
  \setlength{\parskip}{0pt}
  \setlength{\itemsep}{0pt plus 1pt}
	\item According to \citep{ref15}, in the existing non-IID approaches, only one of these, Label skew, Feature, Temporal skew, and Quantity skew, have been discussed, and no research has been done together on both these skews. However, in $\textsc{FedAnil}$, non-IID is addressed from two perspectives: Feature and Label skew.
	\item In the existing approaches to solve the Label and Feature distribution skew, more research has been done from the perspective of Noise-based feature imbalance skew and Synthetic future imbalance skew, and less research has been done on Real-world feature imbalance skew. However, in $\textsc{FedAnil}$, non-IID is addressed from the perspective of Real-world feature imbalance skew.
	\item The existing approaches focused on the part of the attacks, for example, poisoning attacks or inference attacks to protect the training data and client parameters privacy; But {$\textsc{FedAnil}$} focuses on both categories of poisoning and inference attacks and the model privacy and accuracy are well preserved.
\end{itemize}
\vspace{-\topsep}

\begin{table}
  \caption{Comparison between $\textsc{FedAnil}$ and the related work}.
  \setlength{\tabcolsep}{4.0\tabcolsep}
  \centering
   \begin{tabular}{p{4.0cm}|p{1.3cm}|p{1.3cm}|p{1.3cm}}
    \toprule
    \textbf{Ref} & \textbf{robust non-IID} & \textbf{Privacy-Preserving} & \textbf{Secure}\\
    \midrule
    \citep{ref31}&\textcolor{red}{\xmark}&\textcolor{ForestGreen}{\cmark}&\textcolor{ForestGreen}{\cmark}\\
    \citep{ref14}&\textcolor{red}{\xmark}&\textcolor{ForestGreen}{\cmark}&\textcolor{red}{\xmark}\\
    \citep{ref10}&\textcolor{ForestGreen}{\cmark}&\textcolor{ForestGreen}{\cmark}&\textcolor{red}{\xmark}\\
    \citep{ref16}&\textcolor{ForestGreen}{\cmark}&\textcolor{ForestGreen}{\cmark}&\textcolor{red}{\xmark}\\
    \citep{ref2}&\textcolor{ForestGreen}{\cmark}&\textcolor{ForestGreen}{\cmark}&\textcolor{red}{\xmark}\\
    \citep{ref19}&\textcolor{ForestGreen}{\cmark}&\textcolor{ForestGreen}{\cmark}&\textcolor{red}{\xmark}\\
    \citep{ref20}&\textcolor{red}{\xmark}&\textcolor{ForestGreen}{\cmark}&\textcolor{ForestGreen}{\cmark}\\
    \citep{ref35}&\textcolor{red}{\xmark}&\textcolor{ForestGreen}{\cmark}&\textcolor{red}{\xmark}\\
    \citep{ref61}&\textcolor{ForestGreen}{\cmark}&\textcolor{ForestGreen}{\cmark}&\textcolor{red}{\xmark}\\
    \citep{ref62}&\textcolor{ForestGreen}{\cmark}&\textcolor{ForestGreen}{\cmark}&\textcolor{red}{\xmark}\\
    \citep{ref63}&\textcolor{ForestGreen}{\cmark}&\textcolor{ForestGreen}{\cmark}&\textcolor{red}{\xmark}\\
    \citep{ref64}&\textcolor{ForestGreen}{\cmark}&\textcolor{ForestGreen}{\cmark}&\textcolor{red}{\xmark}\\
    \textbf{$\textsc{FedAnil}$}&\textcolor{ForestGreen}{\cmark}&\textcolor{ForestGreen}{\cmark}&\textcolor{ForestGreen}{\cmark}\\
    
    \bottomrule
  \end{tabular}
  \label{tab1}
\end{table}

\section{FedAnil Model}\label{FM1}
\begin{figure}[!t]
\centering
\includegraphics[width=6in]{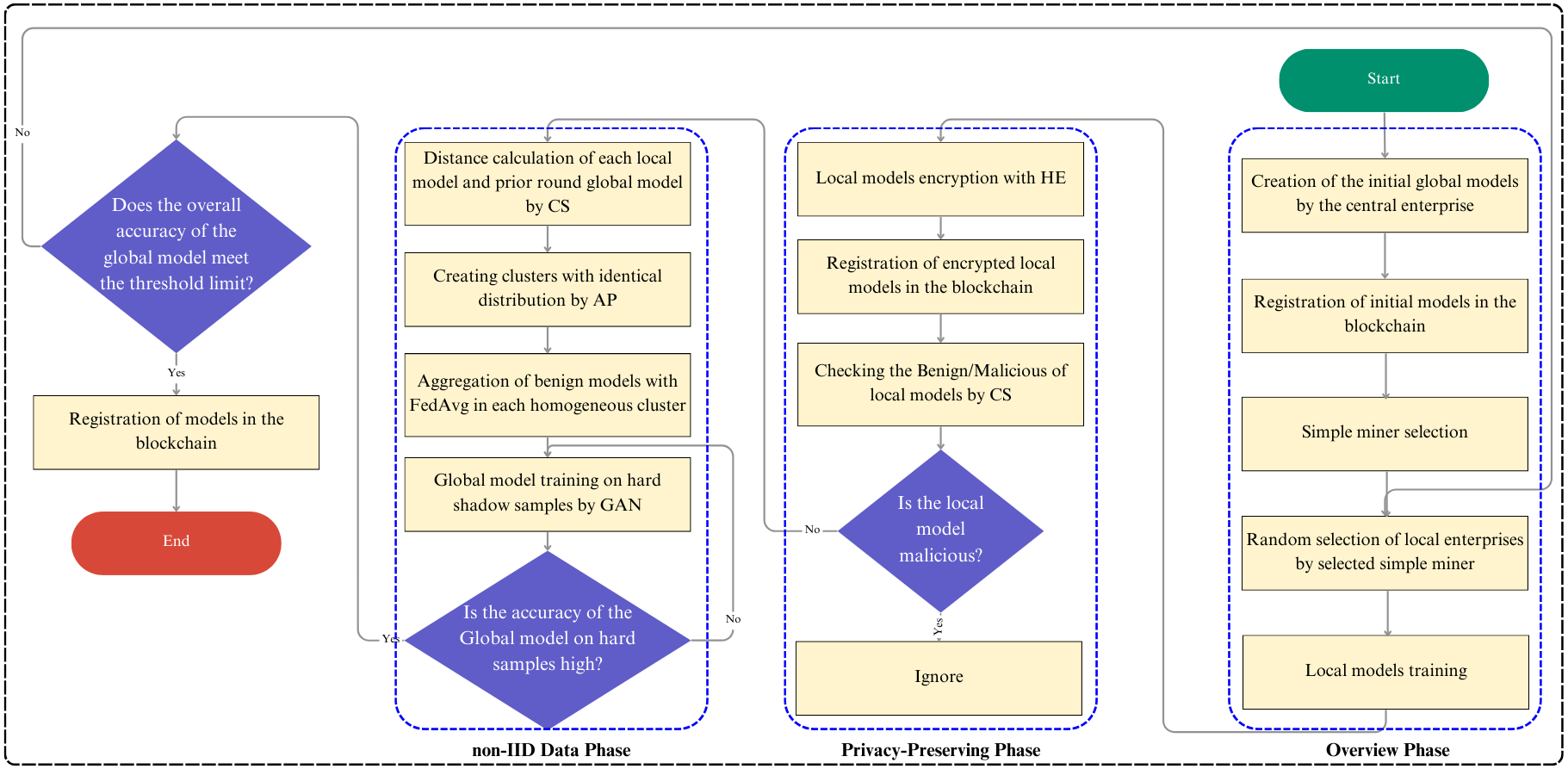}
\caption{In Schematic of $\textsc{FedAnil}$ model.}
\label{fig_FedAnil-Flowchart}
\end{figure}

\begin{figure*}[!t]
	\centering
	{\includegraphics[width=5.0in, frame]{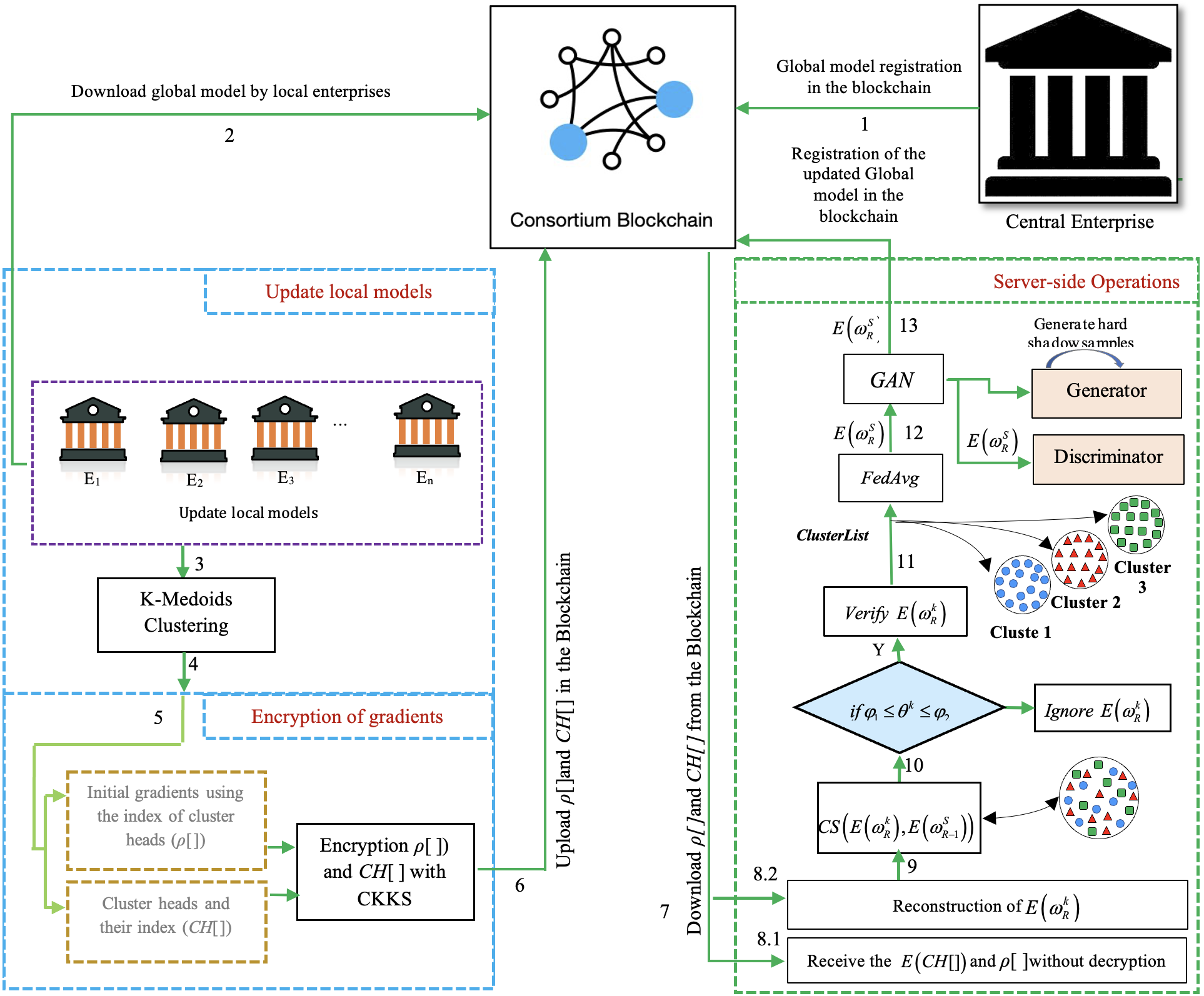}%
		\label{fig_computation_1_case}}
	\hfil
	\caption{The $\textsc{FedAnil}$ model architecture.}
	\label{fig_FedPaC_Model}
\end{figure*}

As shown in Fig. \ref{fig_FedAnil-Flowchart}, the proposed $\textsc{FedAnil}$ model includes three main phases: The Overview, non-IID data, and Privacy-Preserving. The Overview phase explains the necessity of leveraging consortium blockchain and Six steps in a round of global iteration. The details of this phase are brought in Section \ref{FM_Overview}. The non-IID data phase is proposed to address the label and feature distribution skew. The details of this phase will be provided in Section \ref{FM_Unbalanced_non-IID}. Finally, the Privacy-Preserving phase is adopted to address security and privacy concerns against poisoning and inference attacks, which will be explained in detail in Section \ref{FM_Privacy-preserving}. For more clarification, the details of the proposed model phases are shown in Fig. \ref{fig_FedPaC_Model}. 

\subsection{Overview}\label{FM_Overview}
First, in this section, the necessity of leveraging consortium blockchain technology in the $\textsc{FedAnil}$ model is specified, and then six steps in a round of global iteration are explained:
\vspace{-\topsep}
\begin{itemize}
  \setlength{\parskip}{0pt}
  \setlength{\itemsep}{0pt plus 1pt}
 \item Registration of initial models (GloVe, CNN, and ResNet50) in the blockchain to access selected local enterprises.
 \item Initializes the CKKS parameters, specifically registering the public and private keys in the blockchain to each selected local enterprise.
 \item Choosing a simple miner in the blockchain to send the global model to selected local enterprises.
\item Random selection of local enterprises by the selected simple miner.
\item Downloading the global model by selected local enterprises from the blockchain.
\item Upload updated local models to the blockchain by selected local enterprises.
 \item Using blockchain to prevent collude attacks by randomly selecting local enterprises.
 \item To aggregate gradients in a decentralized manner, which addresses the single-point-of-failure problem.
 \item In the consortium blockchain used in the proposed model, a complete system is controlled by a group of enterprises (Consensus Mechanism). They do not allow any decision-making to be performed by the external enterprise.
\end{itemize}
\vspace{-\topsep}

In this paper, three component types are used, namely \emph{miners}, \emph{validators}, and \emph{local enterprises}. All three component types are from the same type of system, i.e., enterprise. Specifically, in $\textsc{FedAnil}$, in each communication round, the role of each component will be randomly according to one of the following three modes:
\vspace{-\topsep}
\begin{itemize}
  \setlength{\parskip}{0pt}
  \setlength{\itemsep}{0pt plus 1pt}
\item It is a \emph{local enterprise} that only updates its local model.
\item It is a \emph{validator} that evaluates the integrity of the received local models on the server side and votes for them. The criterion for evaluating the integrity of local models is based on the equation \eqref{eq_Add_Unbalanced_Non-IID1}. All validators send their votes to the selected simple miner.
\item It is a \emph{miner} that receives the votes of all Validators and determines the integrity (benign/poisoned) of each local model based on the highest vote. Then, it informs the desired local enterprise about the evaluation result of the desired model by creating a new block.
\end{itemize}
\vspace{-\topsep}

The consensus mechanism in the proposed model is the Proof-of-Stake (PoS) mechanism, inspired by the consortium blockchain. Specifically, in each round of blockchain execution, according to the PoS mechanism, the miner with the most stake (reward) is selected among all the miners. Then, the selected simple miner receives the initial global model created by the central enterprise and distributes it to selected enterprises so that local enterprises can update their local models.

In the $\textsc{FedAnil}$, every local enterprise that has updated the benign model is rewarded. On the contrary, if the local enterprise has updated the poisoned model, it will be punished. The validator detects benign/Poisoned models on the server side. Now, any local enterprise with the highest reward stake is selected as a miner. The miner is chosen because it has the highest learning rate among other local enterprises in each communication round.

In $\textsc{FedAnil}$, there are six steps in a round of global iteration. Next, we will explain each step in detail.
\begin{enumerate}
	\item {\bf{\textsc{Initialization.}}} First, the central enterprise creates three initial global models (global vectors for word representation (GloVe), convolutional neural network (CNN), and deep residual networks (ResNet50)). Then, it initializes the \emph{CKKS-FHE} parameters. Specifically, it generates a separate, $p_{k}$, and, $s_{k}$, for each selected enterprise. Afterward, the central enterprise registers both the initial models and the public and private keys in the blockchain so that selected enterprises can receive these initial models and keys.
	\item {\bf{\textsc{Simple miner selection.}}} The role of miners in $\textsc{FedAnil}$ is that they facilitate heavy blockchain operations. In the proposed model, there are two types of miners, namely simple and leader. The role of a simple miner is to evaluate the local model's integrity. In addition, receive the initial global models from the central enterprise and send them to selected local enterprises.
	\item {\bf{\textsc{Enterprises random selection.}}} In this step, the purpose of randomly selecting enterprises is to prevent collude attacks between enterprises. Specifically, in this case, malicious enterprises cannot predict which enterprises will be selected in every round to train the global model due to the random selection among enterprises. Therefore, enterprises will most likely not be able to form collude attacks with possible partner enterprises carrying out collude attacks.
	\item {\bf{\textsc{Leader miner selection.}}}The role of a leader miner is to aggregate the verified local models by the simple miner and update the global model. It should be noted that the leader miner is selected from among the selected simple miners. This means that the selected simple miner has the highest reward.
	\item {\bf{\textsc{Local model training.}}}In this step, the selected enterprises first download the initial models from the blockchain and train them based on their data type in parallel. Second, all local enterprises cluster their vectors with \emph{K-Medoids Clustering}. The clustering output will be two vectors named $\Upsilon$ and the vector $\Psi$. Finally, each enterprise encrypts the vector $\Upsilon$ and the vector $\Psi$ by \emph{CKKS-FHE} according to \eqref{eq_Privacy-preserving3}and then uploads the encrypted gradient vectors to the blockchain.
	\item {\bf{\textsc{Model aggregation and Block generation.}}} In this step, the leader miner (aggregation server) aggregates the verified encrypted local models and updates a global model. Thus, Firstly, local models with similar data distribution are grouped into homogeneous clusters using \emph{CS} and \emph{AP} techniques. Secondly, the FedAvg algorithm is executed on each homogeneous cluster. In the next step, it performs the \emph{WGAN} operation on the updated global model to improve the global model's accuracy by training on $\vartheta$ samples. Finally,  record the final global models in the new block so that the next round of training can be performed by $\Delta c$. This operation will be repeated until the model reaches the desired accuracy. In \eqref{eqModelAgg}, the aggregation process is shown:
	\begin{equation}
		\begin{split}
			E\Bigl(\boldsymbol\omega(\tau)_{r}^{S}\Bigl)=FedAvg\biggl(E\bigl(\boldsymbol\omega(\tau)_{r}^{k=1}\bigl),\cdots, E\bigl(\boldsymbol\omega(\tau)_{r}^{k\in \Delta c}\bigl), p_{k}\biggl).
			\label{eqModelAgg}
		\end{split}
	\end{equation}
\end{enumerate} 

In \eqref{eqModelAgg}:
\vspace{-\topsep}
\begin{itemize}
\setlength{\parskip}{0pt}
  \setlength{\itemsep}{0pt plus 1pt}
	\item \emph{$E\bigl(\boldsymbol\omega(\tau)_{r}^{S}\bigl)$}: The encrypted $\boldsymbol\omega(\tau)_{r}^{S}$.
	\item \emph{$E\bigl(\boldsymbol\omega(\tau)_{r}^{k}\bigl)$}: The encrypted $\boldsymbol\omega(\tau)_{r}^{k}$.
	\item \emph{$p_{k}$}: The public key. 
\end{itemize}
\vspace{-\topsep}

\subsection{Addressing the non-IID}\label{FM_Unbalanced_non-IID}
In the proposed model and on the central enterprise side, an approach based on \emph{CS}, \emph{AP}, and \emph{WGAN} is adopted to ensure the reliability and robustness of the $\textsc{FedAnil}$ model against the label and feature distribution skew. Therefore, this subsection addresses the challenges of label and feature distribution skew. The summary details of this approach are given below:

\vspace{-\topsep}
\begin{itemize}
\setlength{\parskip}{0pt}
  \setlength{\itemsep}{0pt plus 1pt}
    \item {\bf{Clustering heterogeneous models:}} On the server side, a clustering technique based on \emph{CS} and \emph{AP} is used. The \emph{CS} technique separates local models by type. Then, using the \emph{AP} clustering technique, models of the same type with close and similar data distributions are grouped in a homogeneous cluster. Specifically, using the \emph{CS} technique in \eqref{eq_Add_Unbalanced_Non-IID1}, the distance between the $\boldsymbol\omega(\tau)_{{r}}^{k}$ and $\boldsymbol\omega(\tau)_{{r-1}}^{S}$is calculated and stored in a list, then using the \emph{AP} algorithm in \eqref{eq_Add_Unbalanced_Non-IID2}, the cluster members are specified. After performing the clustering operation, the aggregation process can be performed correctly on each homogeneous cluster by {$\textsc{FedAvg}$}without reducing the accuracy of the global model.
    \begin{equation}
\begin{split} \emph{ClusterList[1$\dots$$\mathcal{M}$] = $AP\Bigl(\theta(\tau)_{r}^{k=1},\dots,\theta(\tau)_{r}^{k\in \Delta c}\Bigl)$};
\label{eq_Add_Unbalanced_Non-IID2}
\end{split}
\end{equation}

In \eqref{eq_Add_Unbalanced_Non-IID2}:
\begin{itemize}
\item \emph{ClusterList[1$\dots$$\mathcal{M}$]}: List of clusters created by \emph{AP} and \emph{CS}.
\item $\theta(\tau)_{r}^{k}$: Distance between $E\bigl(\boldsymbol\omega(\tau)_{{r}}^{k}\bigl)$ and $E\bigl(\boldsymbol\omega(\tau)_{{r-1}}^{S}\bigl)$.
\end{itemize}
    \item {\bf{Dealing with label and feature distribution skew:}} After the clustering operation and creating homogeneous global models, \emph{WGAN} has dealt with the label and feature distribution skew. Therefore, all three types of global models are trained by \emph{WGAN} with Hard shadow samples. In \emph{WGAN}, a generator and a discriminator (global model) try to learn Hard shadow samples. Specifically, for adversarial training, a generator produces shadow samples containing Hard samples similar to the Hard samples of local enterprises. On the other hand, the discriminator in this adversarial training is considered the global model. In fact, in this adversarial model, there is competition between the generator and the global model in each round of the \emph{FL} process. In this competition, the goal is for the generator to generate so many Hard shadow samples that the global model (detector) reaches a step where it can learn all of this Hard shadow data during this adversarial training process. For example, we can consider samples from the FEMNIST dataset in different local enterprises with feature and label skew. Therefore, it is possible to detect the bias on the "italic" feature in the handwritten numbers, which led to the misclassification of the class label, by using adversarial training between the generator and the global model so that the correct classification is done. As a result, in this case, the global model can preserve accuracy and convergence on local models that follow their local training through this global model.
    \item {\bf{Fending off poisoning attacks:}} Data poisoning, model poisoning, and collude attacks have been prevented to ensure robustness and global model accuracy. Full details of this section are given in Section \ref{FM_Privacy-preserving} ({\bf{\textsc{Step 1}}}, {\bf{\textsc{Step 2}}}, and {\bf{\textsc{Step 3}}}).
\end{itemize}

The further details of addressing label and feature distribution skew are given below:

{\bf{\textsc{Label and Feature distribution skew.}}} In the existing approaches to solve the label and feature distribution skew challenge, more research has been done on the aspects of noise-based feature imbalance skew and synthetic future imbalance skew, and less research has been done on real-world feature imbalance skew \citep{ref15}. $\textsc{FedAnil}$ addresses the real-world feature imbalance skew aspect. This step proposes a \emph{WGAN-based} \citep{ref37} approach to solve the label and feature distribution skew challenge to improve the global model accuracy and convergence.

\emph{WGAN} is an alternative to traditional \emph{GAN} training. Because the traditional \emph{GAN} suffers from the challenge of poor sample diversity and difficult training, \emph{WGAN} is a suitable alternative to solve both the challenges of poor sample diversity and difficult training. Since, in $\textsc{FedAnil}$, the server cannot directly access the private training data of enterprises. Therefore, it creates a challenge for accurate prediction. In $\textsc{FedAnil}$, high-quality and diverse samples are needed to estimate the overall distribution of enterprises' data without violating data privacy. This goal is satisfied by \emph{WGAN}. Specifically, this step simulates the latent space of local models through a $\mathbb{G}$ by producing $\vartheta$ and uses it to train $\mathbb{D}$ to learn $\vartheta$ of local models. A $\mathbb{G}$ produces $\vartheta$ similar to $\vartheta$ of local enterprises; on the other hand, the $\mathbb{D}$ is considered the global model. Each round has competition between $\mathbb{G}$ and $\mathbb{D}$. In this competition, the goal is for $\mathbb{G}$ to produce so many $\vartheta$ that $\mathbb{D}$ reaches a step where it can learn all these $\vartheta$ during this adversarial training process. According to \eqref{eq_LabelFeature_skew}, the server creates a $\mathbb{G}$ that generates $\vartheta$ to gain the local enterprise's data distribution:
\begin{equation}
	\begin{split} \emph{$\vartheta$ = $\mathbb{G}\Bigl(\mathbb{Z},\mathbb{Y},\Theta \Bigl)$},
		\label{eq_LabelFeature_skew}
	\end{split}
\end{equation}

where Gaussian noise is denoted, $\mathbb{Z}$ $\sim$ $\mathbb{N}$$\bigl($0$, $1$\bigl)$, the class label $\vartheta$ is denoting $\mathbb{Y}$, sampled from the predefined distribution $p_t(\mathbb{Y})$, and finally, $\Theta$ is the $\mathbb{G}$ parameter. Therefore, according to \eqref{eq_LabelFeature_skew}, the goal is to solve \eqref{eq_LabelFeature_skew2}:
\begin{equation} 
	\begin{split}
		\underset {E(\boldsymbol\omega^S)} \min\hspace{0.06cm}{\mathbb{E}}_{\substack{\mathbb{Z} \sim \mathbb{N}\bigl(0,1\bigl),\\ \mathbb{Y} \sim p_t\bigl(\mathbb{Y}\bigl)}} \bigr[ \mathcal{L}_{_{LG}} \bigr]=\underset {E(\boldsymbol\omega^S)} \min\hspace{0.06cm}{\mathbb{E}}_{\substack{\mathbb{Z} \sim \mathbb{N}\bigl(0,1\bigl),\\ \mathbb{Y} \sim p_t\bigl(\mathbb{Y}\bigl)}} \Biggr[\sum _{k \in \Delta c} \mathcal{L}_{_{LG}}^{k} \Biggr],
		\label{eq_LabelFeature_skew2}
	\end{split}
\end{equation}

where $\mathcal{L}_{_{LG}}^{k}$ is the model difference between the $E(\boldsymbol\omega^{S})$ and the $E(\boldsymbol\omega^{k})$.
\begin{equation}
	\begin{split}
		\mathcal{L}_{_{LG}}^{k}=D_{KL}\biggl(\sigma\Bigl(D\bigl(\vartheta;E(\boldsymbol\omega^S)\bigl)\biggl)\Vert\sigma\biggl(D\bigl(\vartheta;E(\boldsymbol\omega^k)\bigl)\Bigl)\biggl),
		\label{eq_LabelFeature_skew3}
	\end{split}
\end{equation}

where the $\textsc{Kullback-Leibler}$ divergence is represented by the $D_{KL}$. $\sigma$ represents the Softmax function, which obtains the prediction score $\vartheta$. The classifier is marked with $D$. Therefore, this step in $\textsc{FedAnil}$ follows two goals that are formulated in \eqref{eq_LabelFeature_skew4}:

\begin{enumerate}
	\item Producing $\vartheta$ by generator $\mathbb{G}$ to increase $\mathcal{L}_{_{LG}}$.
	\item Training the global model on $\vartheta$ to reduce $\mathcal{L}_{_{LG}}$.
\end{enumerate}
\begin{equation} 
	\begin{split}
		\underset {E(\boldsymbol\omega^S)} \min\hspace{0.06cm} \underset {\Theta} \max\hspace{0.06cm}{\mathbb{E}}_{\substack{\mathbb{Z} \sim \mathbb{N}\bigl(0,1\bigl),\\ \mathbb{Y} \sim p_t\bigl(\mathbb{Y}\bigl)}} \bigr[ \mathcal{L}_{_{LG}} \bigr],
		\label{eq_LabelFeature_skew4}
	\end{split}
\end{equation}

By reducing the value of the $\mathcal{L}_{_{LG}}$ parameter, the classifier error is reduced, and the $\boldsymbol\omega(\tau)_{r}^{S}$ accuracy is closer to the $\boldsymbol\omega(\tau)_{r}^{k}$ accuracy, and this is the global model convergence. Because $\textsc{FedAnil}$ randomly retrieves part of the training data enterprises on the server, it may violate the privacy of local models in FL. However, $\vartheta$ only gains the training data with high-quality feature patterns humans cannot understand. Moreover, because all parameters of local models are processed by \emph{CKKS-FHE} in encrypted form, the local model's privacy is preserved. The detailed addressing of the non-IID data procedure is represented in Algorithm \ref{alg_Unbalanced_and_Non-IID}.

Therefore, in the scenarios where the data distribution is uneven (non-IID), performing these steps ({\emph{Clustering heterogeneous models}}, {\emph{Dealing with label and feature distribution skew}}, and {\emph{Fending off poisoning attacks}})gives the reliability that the global model does not suffer from the challenges of feature and label skew, and with avoid poisoning attacks, preserve the overall robustness of the model.

\subsection{Addressing privacy-preserving}\label{FM_Privacy-preserving}
\begin{figure}[!t]
\centering
\includegraphics[width=3.5in]{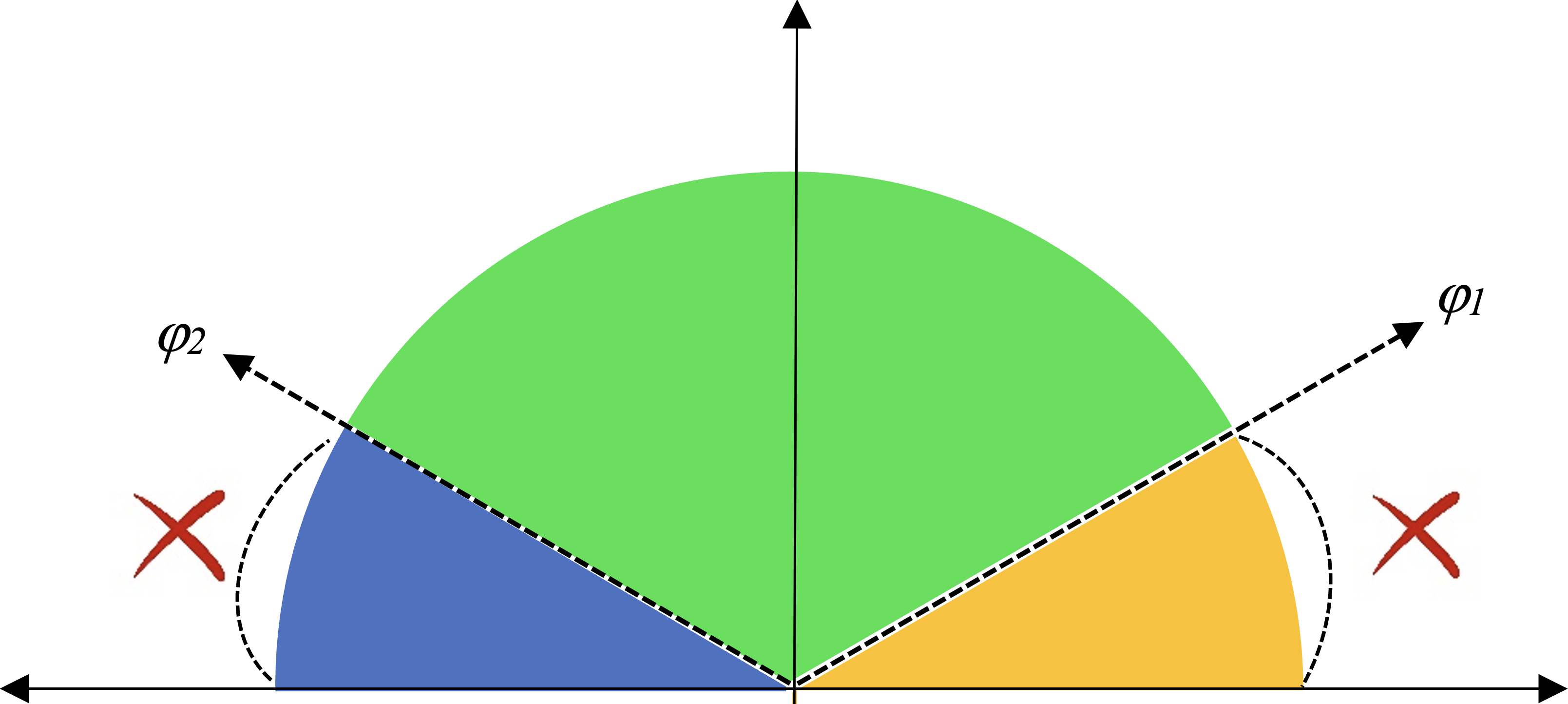}
\caption{$\varphi_1$ and $\varphi_2$ in the $\textsc{FedAnil}$ model}.
\label{fig_Fending_off_Data_poisoning}
\end{figure}

This subsection focuses on reducing data security and privacy concerns and improving the global model's accuracy by preventing attacks like Poisoning and Inference. It should be noted that the exchange of the global and local models is done only through the Consortium Blockchain network and in a private and distributed manner. Therefore, in the $\textsc{FedAnil}$ to reduce data security and privacy concerns in enterprises, three steps are considered as follows:

{\bf{\textsc{Step 1: Fending off the Data poisoning attack.}}} According to \eqref{eq_Add_Unbalanced_Non-IID1}, on the server side, after computing the \emph{CS} between $E\bigl(\boldsymbol\omega(\tau)_{{r-1}}^{S}\bigl)$, and, $E\bigl(\boldsymbol\omega(\tau)_{{r}}^{k}\bigl)$, the state of the local models is checked with a condition. The purpose of this step is to detect and prevent data poisoning attacks. More precisely, a data poisoning attack can use the revealed statistical distribution of data to perform a poisoning attack. In this step, enterprises whose updated gradient vector was outside the range of two thresholds in the proposed model are ignored. Also, if this value is not between $\varphi_1$ and $\varphi_2$ (green range in Fig. \ref{fig_Fending_off_Data_poisoning}) in $5$ rounds, the enterprise is considered a malicious enterprise. Moreover, its data is considered poison data and is excluded from participating in the federal learning process.
\begin{equation}
\begin{split}
\theta/CS \Bigl(E\Bigl(\boldsymbol\omega(\tau)_{{r}}^{k}\Bigl), E\Bigl(\boldsymbol\omega(\tau)_{{r-1}}^{S}\Bigl)\Bigl)=
\frac{\Bigl \langle \Delta E\Bigl(\boldsymbol\omega(\tau)_{{r}}^{k}\Bigl), \Delta E\Bigl(\boldsymbol\omega(\tau)_{{r-1}}^{S}\Bigl) \Bigr \rangle}{\Big\| \Delta E\Bigl(\boldsymbol\omega(\tau)_{{r}}^{k}\Bigl) \Big\|*\Big\| \Delta E\Bigl(\boldsymbol\omega(\tau)_{{r-1}}^{S}\Bigl) \Big\|}.
\label{eq_Add_Unbalanced_Non-IID1}
\end{split}
\end{equation}

In \eqref{eq_Add_Unbalanced_Non-IID1}:
\vspace{-\topsep}
\begin{itemize}
\setlength{\parskip}{0pt}
  \setlength{\itemsep}{0pt plus 1pt}
\item $\theta$/\emph{CS}: The angle between the two vectors. 
\item $\Delta E\bigl(\boldsymbol\omega(\tau)_{{r}}^{k}\bigl)$: The enterprise \emph{k} gradients in the round \emph{r}.
\item $\Delta E\bigl(\boldsymbol\omega(\tau)_{{r-1}}^{S}\bigl)$: The server gradients in the prior round.
\end{itemize}
\vspace{-\topsep}

The details of Fig. \ref{fig_Fending_off_Data_poisoning} are as follows. This figure includes the interval of changes between $0$ and $180$ degrees, which changes the numerical value of the cosine between a continuous interval of $+1$ and $-1$. This figure has three ranges marked from right to left with yellow, green, and blue colors, and the ranges of these three colors are separated by using two threshold ranges, $\varphi_1$ and $\varphi_2$.

In Fig. \ref{fig_Fending_off_Data_poisoning}:
\vspace{-\topsep}
\begin{itemize}
  \setlength{\parskip}{0pt}
  \setlength{\itemsep}{0pt plus 1pt}
\item {\bf{\textsc{Yellow color range.}}} In this range, local enterprises may not have been able to provide a benign update due to the noise injection made by data poisoning attacks on their training data. Therefore, partial updates are removed due to the small cosine angle formed between two vectors. 
\item {\bf{\textsc{Blue color range.}}} By adding noise to enterprises' training data, intruders try to make enterprises behave dissimilarly to the global model, and this will cause a large angle (low similarity) between these two gradient vectors. As a result, to prevent data poisoning attacks, the local gradient vector of enterprises that fall within this range is also prevented from continuing the process of participating in model aggregation. 
\item {\bf{\textsc{Green color range.}}} Finally, the local gradient vectors, $E\bigl(\boldsymbol\omega(\tau)_{{r}}^{k}\bigl)$, that makes an angle, $\theta(\tau) _{r}^{k}$, with the global gradient vector, $E\bigl(\boldsymbol\omega(\tau)_{{r-1}}^{S}\bigl)$, in the green range, i.e. between $\varphi_1$ and $\varphi_2$, have a healthy and noise-free local gradient vector. When a locally updated gradient vector is determined to be safe by the \emph{CS} technique, this vector can be sent to subsequent steps. 
\end{itemize}
\vspace{-\topsep}

{\bf{\textsc{Step 2: Fending off the Collude attack.}}} Preventing the collude attack is based on two techniques: Consortium blockchain and random selection in the $\textsc{FedAnil}$ model. In this model, because, $E\bigl(\boldsymbol\omega(\tau)^{S}\bigl)$, is an average of the behavior of, $E\bigl(\boldsymbol\omega(\tau)_{{r}}^{k}\bigl)\in \Delta c$\hspace{0.12cm}, selected enterprises must change their behavior to be able to influence the global model with their attack, which is called a collude attack. Since the global model is an unchangeable reference and is directly manipulated by enterprises, no local enterprises can bias this model in their favor. Therefore, it is predicted that this process will protect enterprises from colluding attacks to a very good extent. On the other hand, the $\textsc{FedAnil}$ does not consider all the enterprises to receive the update of the global model, and the simple miner makes a random choice among the enterprises, collude attacks can be prevented to a very good extent by randomly choosing the enterprises. Unless a collude attack occurs on a large scale between enterprises, this is also ruled out because the participating enterprises work in a consortium blockchain structure restricted to the public. 

{\bf{\textsc{Step 3: Fending off the Model poisoning, Membership inference, and Reconstruction attacks.}}} In the $\textsc{FedAnil}$ model, the fusion of consortium blockchain and \emph{CKKS-FHE} is employed to prevent these attacks. In addition, $\textsc{FedAnil}$ is presumed that the server is defined as $\textsc{honest-but-curious}$. 

A model poisoning attack is an attack that aims to change and manipulate local parameters. Therefore, $\textsc{FedAnil}$ should be able to prevent the disclosure of updated parameters during the aggregating process and update the global model. As a result, in $\textsc{FedAnil}$, relying on the \emph{CKKS-FHE} can prevent the disclosure of local parameters and allow servers to perform the necessary computation to aggregate and create a global model. In a membership inference attack, the adversary inspects the global model to determine whether the desired sample exists in the training dataset. The attacker has the architecture used to train the real data in this attack. Hence, it can infer real training data using fake datasets and updated parameters so that privacy violations occur in enterprises. Therefore, according to \eqref{eq_Privacy-preserving3}, to avoid revealing the parameters, both of the vectors $\Psi$ and $\Upsilon$ are encrypted by \emph{CKKS-FHE}, and finally, both vectors are recorded in the blockchain.

\begin{equation}
{{E\Bigl(CH[\hspace{0.10cm}], \rho[\hspace{0.10cm}]\Bigl)}} \longleftarrow Encrypt\Bigl(CH[\hspace{0.10cm}], \rho[\hspace{0.10cm}], p_{k} \Bigl). \\
\label{eq_Privacy-preserving3}
\end{equation}

In \eqref{eq_Privacy-preserving3}:
\vspace{-\topsep}
\begin{itemize}
  \setlength{\parskip}{0pt}
  \setlength{\itemsep}{0pt plus 1pt}
\item $\rho[\hspace{0.10cm}]$, $En\bigl(\rho[\hspace{0.10cm}]\bigl)$: $\Upsilon$, Vector of encrypted $\Upsilon$.
\item $p_{k}$: The public key.
\item $CH[\hspace{0.10cm}]$, $E\bigl(CH[\hspace{0.10cm}]\bigl)$: $\Psi$, Vector of encrypted $\Psi$.
\end{itemize}
\vspace{-\topsep}

In the reconstruction attack, the honest-but-curious server performs the attack by accessing the updated local enterprise parameters. In the proposed model to defend against this attack, \emph{CKKS-FHE} is leveraged to encrypt the local parameters so attackers cannot access a reference parameter set to adjust model parameters. As a result, when the intruder does not have control over the encrypted models, it cannot compare the output obtained from the fake data with any authority to form a privacy violation in the enterprise data by setting these parameters. The detailed addressing of the privacy-preserving and the involved fending off the poisoning and inference attacks procedure is represented in the Algorithm \ref{alg_Privacy-preserving}.

\section{Convergence analysis of FedAnil Model}\label{CA1}
{In this subsection, First some cases in which the proposed $\textsc{FedAnil}$ model may oscillate and lead to model divergence are listed, Next, convergence analysis is discussed in detail.}

\vspace{-\topsep}
\begin{itemize}
  \setlength{\parskip}{0pt}
  \setlength{\itemsep}{0pt plus 1pt}
\item In $\textsc{FedAnil}$, due to local enterprises’ non-IID datasets, the weight parameter directions may diverge when trained to fit models of different data distributions, even when initialized with identical weights.
\item The local class imbalance (non-IID) will cause weight divergence and accuracy loss. The proposed model may diverge, especially when the amount of training data or classes among local enterprises is unbalanced.
\item In another case of non-IID data challenges such as:
\begin{itemize}
  \setlength{\parskip}{0pt}
  \setlength{\itemsep}{0pt plus 1pt}
    \item Data type skew
    \item Cluster skew
\end{itemize}
\item More differences between Local and Global logit distribution.
\item Intuitively, large \emph{epochs} may cause local models to drift too far away from the initial starting point, leading to potential divergence.
\item In $\textsc{FedAnil}$, larger \emph{batch sizes} work well but can result in divergence in $\textsc{FedAnil}$ implementations.
\item Model overfitting and the poisoning attack scenarios
\end{itemize}

Finally, although the $\textsc{FedAnil}$ model has proposed a solution for addressing non-IID data and poisoning attacks, the probability of occurrence of these divergences is very low but possible. Also, \emph{epochs} and \emph{batch sizes} hyperparameters are fine-tuned during execution, but model divergence is possible in large values. Moreover, some divergence cases are defined as future studies.

For the convergence analysis, assuming that $F(\boldsymbol\omega)$ is non-convex, we discuss the convergence of the $\textsc{FedAnil}$ model. Our analysis has the following three assumptions:

	{\bf{\textsc{Assumption 1}}} {$\textsc{(Bounded Divergence)}$}. Assume that $\parallel$$\nabla{F^k(\boldsymbol\omega)}$$\parallel$$^2$$\leq C$. The divergence of the $\boldsymbol\omega^{k}$ model of each enterprise from the $\overline{\boldsymbol\omega}_{{r}}$ average model will be according to $\Vert \boldsymbol\omega_{{r}}^{k}-\overline{\boldsymbol\omega}_{{r}}\Vert^2\leq \delta^2\overline{r}^{2}C$ (for some constant $C < \infty$ and all $k$). Therefore, reducing the $\overline{r}$ variable causes the {\textsc{FedAvg}} algorithm to behave like standard {\textsc{Gradient Descent}}. However, if the variable $\overline{r}$ increases, the communication and the convergence rate in the {\textsc{FedAvg}} algorithm will decrease.

\begin{algorithm}
\caption{non-IID Data in the $\textsc{FedAnil}$}
\begin{algorithmic}[1]
\Procedure{non-IID}{}{\color{blue}\Comment{\emph{Sect.\hspace{0.03cm}IV(B),\hspace{0.03cm}Step\hspace{0.03cm}1,\hspace{0.05cm}2}}}
\State {\bf{\textsc{Label\hspace{0.03cm}and\hspace{0.05cm}Feature\hspace{0.06cm}Skew();}}}
            \State $\mathcal{M}=[CNN, ResNet50, GloVe$];
            \For {{$each$ $\boldsymbol\omega{{(\tau)}}^{k}$} {$\in$$\mathcal{M}$}\hspace{0.12cm} in parallel}
            \For {{$r=1$ $to$ $R$}}
            \For {{$each$ $k$} {$\in$  $\Delta c$}\hspace{0.12cm}in parallel}
            \State $\theta(\tau)_{r}^{k} = CS\Bigl(E\Bigl(\boldsymbol\omega(\tau)_{{r}}^{k}\Bigl), E\Bigl(\boldsymbol\omega(\tau)_{{r-1}}^{S}\Bigl)\Bigl)$;
            \State \emph{ClusterList[1$\dots$$\mathcal{M}$]=$AP\Bigl(\theta(\tau){r}^{k=1},\theta(\tau)_{r}^{k=2},\dots,\theta(\tau)_{r}^{k\in \Delta c}\Bigl)$};            
            \State $E\Bigl(\boldsymbol\omega(\tau)_{r}^{S}\Bigl)\gets FedAvg(\emph{ClusterList[1$\dots$$\mathcal{M}$])}$;
            \For {{$\tau=1$ $to$ $\mathcal{M}$}}
            \State $\mathbb{D} \longleftarrow E\Bigl(\boldsymbol\omega(\tau)_{r}^{S}\Bigl)$;
            \While{$E\Bigl(\boldsymbol\omega(\tau)_{r}^{S}\Bigl)$} didn't\hspace{0.06cm}detect\hspace{0.06cm}$\vartheta$
            \State ${\mathbb{G}\overset{generate}{\Longrightarrow} \emph{$\vartheta$}}$ according to \eqref{eq_LabelFeature_skew}; 
            \State{Train $E\Bigl(\boldsymbol\omega(\tau)_{r}^{S}\Bigl)$} with $\vartheta$;  
            \State{Calculate $\mathcal{L}_{_{LG}}^{k}$} according to \eqref{eq_LabelFeature_skew3}
            \If{$\mathcal{L}_{_{LG}}^{k} \leq \varphi$}
            \State{Break;}
            \EndIf
            \EndWhile
        \EndFor
        \EndFor
        \EndFor
        \EndFor
\EndProcedure
\end{algorithmic}
\label{alg_Unbalanced_and_Non-IID}
\end{algorithm}

{\bf{\textsc{Assumption 2}}} {$\textsc{($\beta$-smoothness)}$}. Assuming $\nabla F^{k}(\boldsymbol\omega)$ is $\beta$ smoothness, therefore, $\Vert (\nabla F^{k}(\boldsymbol\omega) - \nabla F(\boldsymbol\omega _{\ast}))\Vert \leq \beta \Vert (\boldsymbol\omega^{k} - \boldsymbol\omega _{\ast}) \Vert$, where $\forall$$\boldsymbol\omega^{k}$, $\boldsymbol\omega _{\ast}$ $\in$ $\mathbb{R}^d$ \citep{ref39}. Where $\beta$ is a positive constant.

{\bf{\textsc{Assumption 3.}}} Assuming $F^{k}(\boldsymbol\omega)$ in selected local enterprises be locally convex. Hence, the following formula will hold for $F^{k}(\boldsymbol\omega)$. $F^{k}[(\wp\boldsymbol\omega+(1-\wp)\boldsymbol\omega _{\ast})] \leq [\wp F^{k}(\boldsymbol\omega)+(1-\wp)F(\boldsymbol\omega _{\ast})]$, $\forall$$\boldsymbol\omega$, $\boldsymbol\omega _{\ast}$ $\in$ $\mathbb{R}^d$, $\wp \in [0,1]$ and $\wp$ is a positive constant, and distance for both $\boldsymbol\omega$ and $\boldsymbol\omega _{\ast}$ at a radius $ri > 0$. In the next discussion, we prove the convergence of the $\textsc{FedAnil}$ model weight parameter $\boldsymbol\omega^{k}$ in training local models.

\begin{algorithm}
\small
	\caption{Privacy-Preserving in the $\textsc{FedAnil}$}
	\begin{algorithmic}[1]
		\Procedure{Privacy-Preserving}{}
		\State {\bf{\textsc{Poisoning Attacks();}}}
		{\color{blue} \Comment{\emph{Sect.\hspace{0.03cm}IV(C),\hspace{0.03cm}Step\hspace{0.03cm}1,\hspace{0.05cm}2}}}
		\For {{$each$ $k$} {$\in$  $\Delta c$}\hspace{0.12cm}in parallel}
		\State $\varphi_1$ = $-0.7$
		\State $\varphi_2$ = $+0.7$
		\State $\theta(\tau) ^{k} = CS\Bigl(E\Bigl(\boldsymbol\omega(\tau)_{{r}}^{k}\Bigl), E\Bigl(\boldsymbol\omega(\tau)_{{r-1}}^{S}\Bigl)\Bigl)$;        
		\If{$\varphi_1$ $\leq$ $\theta(\tau)^{k}$ $\leq$ $\varphi_2$}
		\State $Verify$ $E\Bigl(\boldsymbol\omega(\tau)_{{r}}^{k}\Bigl)$;
		\Else
		\State $Ignore$ $E\Bigl(\boldsymbol\omega(\tau)_{{r}}^{k}\Bigl)$;
		\State $\chi^k++$;
		\EndIf
		\If{$\chi^k=5$}
		\State $Discard$ $E\Bigl(\boldsymbol\omega(\tau) ^{k}\Bigl)$;
		\EndIf
		\EndFor
		\State {\bf{\textsc{Inference Attacks();}}} {\color{blue} \Comment{\emph{Sect.\hspace{0.03cm}IV(C),\hspace{0.03cm}Step\hspace{0.03cm}3}}}
		\For {{$each$ $k$} {$\in$  $\Delta c$}\hspace{0.12cm}in parallel}
		\State {
			$Encryption$ $\Bigl(\rho[\hspace{0.10cm}], CH[\hspace{0.10cm}], CKKS \Bigl)$;
		}  
		\EndFor
		\EndProcedure
	\end{algorithmic}
	\label{alg_Privacy-preserving}
\end{algorithm}
{\bf{\textsc{Theorem 1.}}} For $\beta$ as a constant, if $\eta\leq \frac{1}{\beta}$, then there exists $\Vert (F^{S}(\boldsymbol\omega_{r+1}) - F^{S}(\boldsymbol\omega _{\ast}))\Vert \leq \Vert (F^{S}(\boldsymbol\omega_{r}) - F^{S}(\boldsymbol\omega _{\ast})) \Vert$, where $F^{S}(\boldsymbol\omega_{r})$ denotes the global aggregated model loss function. On the server side, $\boldsymbol\omega_{r}$ and $\boldsymbol\omega _{\ast}$ indicate the normal model and the optimized model at the round $r$.

{\bf{\textsc{Proof.}}} Following the discussion, the proof of the $\Vert \bigl(F^{S}(\boldsymbol\omega_{r}) - F^{S}(\boldsymbol\omega _{\ast})\bigl)\Vert^2$ is given. $\Vert \bigl(F^{S}(\boldsymbol\omega_{r-1}) - \eta\nabla F^{S}(\boldsymbol\omega_{r-1}) - F^{S}(\boldsymbol\omega _{\ast})\bigl)\Vert^2$ = $\Vert \bigl(F^{S}(\boldsymbol\omega_{r-1}) - F^{S}(\boldsymbol\omega _{\ast})\bigl)\Vert^2 - \bigl(2\eta\nabla (F^{S}(\boldsymbol\omega_{r-1})^{\mathbb{GI}}F^{S}(\boldsymbol\omega_{r-1})\bigl) - \bigl(F^{S}(\boldsymbol\omega _{\ast})) + \eta^2\Vert\nabla F^{S}(\boldsymbol\omega_{r-1})\bigl)\Vert^2 \leq \Vert \bigl(F^{S}(\boldsymbol\omega_{r-1}) - F^{S}(\boldsymbol\omega _{\ast})\Vert^2 - \eta\frac{\Vert\nabla\boldsymbol\omega_{r-1})\Vert^2}{\beta} + \eta^2\Vert\nabla F^{S}(\boldsymbol\omega_{r-1})\bigl)\Vert^2 = \Vert \bigl(F^{S}(\boldsymbol\omega_{r-1}) - F^{S}(\boldsymbol\omega _{\ast}))\bigl)\Vert^2 - \eta(\frac{1}{\beta} - \eta)\Vert\nabla F^{S}(\boldsymbol\omega_{r-1})\Vert^2$. The proof of {{\textsc{Theorem 1}}} is finished.

In the end, the following relation is output: $\Vert F^{S}(\boldsymbol\omega_{r}) - F^{S}(\boldsymbol\omega _{\ast})\Vert^2 \leq \Vert F^{S}(\boldsymbol\omega_{r-1}) - F^{S}(\boldsymbol\omega _{\ast})\Vert^2$.

{\bf{\textsc{Corollary 1}}} {$\textsc{(anti-inference and anti-poisoning)}$}. Suppose the number of benign enterprises exceeds the number of intruder enterprises. In that case, the proposed anti-inference and anti-poisoning converge to a model that is the same as benign enterprise models.

{\bf{\textsc{Proof.}}} To prove the above Corollary, allow us to clarify with specific scenarios. In the scenario $1$, all local enterprises are benign. In the scenario $2$, all local enterprises are intruders. Intruder enterprises launch poisoning and inference attacks to destroy model accuracy and privacy. The {$\textsc{FedAnil}$} model will converge in both scenario $1$ and scenario $2$. However, the direction of the final model obtained from Scenario $1$ and Scenario $2$ will differ. The global model for scenarios $1$ and $2$ are marked with $\boldsymbol\omega^{S(benign)}$ and $\boldsymbol\omega^{S(intruder)}$ respectively. The final model obtained after the last round will be a model between $\boldsymbol\omega^{S(benign)}$ and $\boldsymbol\omega^{S(intruder)}$. If the $\mu$ variable is the percentage of intruder enterprises, the final aggregated model after the $r^{th}$ round is denoted as \eqref{eqCorollary1}:
\begin{equation}
\boldsymbol\omega^{S}=\Bigr[(1-\mu) \ast \boldsymbol\omega_{r}^{S(benign)}\Bigr]+\Bigr[\mu \ast \boldsymbol\omega_{r}^{S(intruder)}\Bigr].
\label{eqCorollary1}
\end{equation}

According to\eqref{eqCorollary1}, if most of the local enterprises are benign (i.e. $\mu \approx 0.2$), then the final global model will be near to $\boldsymbol\omega^{S(benign)}$. Here, $\mu$ will positively bring the aggregated model closer to the benign enterprises model, and the proposed anti-inference and anti-poisoning-based aggregated global model decreases the effect of intruder enterprises in each round. The proof of {\textsc{Corollary 1}} is finished.

Next, we explain the convergence of aggregated local models with the same distribution. It should be noted that in the{$\textsc{FedAnil}$} model based on FL, there are $n$ local enterprises. The dataset of enterprises is represented by $DS_1, DS_2,\ldots, DS_n$, and each has a different data distribution $p^\varrho(\varrho=1,2,\ldots,n)$. Assuming that stochastic gradients $SG(.)$ are unbiased with a different probability distribution in each round. i.e, $\mathbb{E}[SG^\varrho(\boldsymbol\omega_{r})]=\nabla F^\varrho(\boldsymbol\omega_{r})$.

{\bf{\textsc{Theorem 2.}}} In the {$\textsc{FedAnil}$} model, after solving the non-IID challenge (discussed in Section \ref{FM_Unbalanced_non-IID}), it is assumed that the set of uploaded weight parameters is selected which have datasets with the same distribution $p^\varrho$. We will have the following relation, compared to the {$\textsc{FedAvg}$}: 
\begin{equation}
\mathbb{E}\Vert \boldsymbol\omega_{r}^{\varrho} - \boldsymbol\omega _{\ast}^{\varrho} \Vert^2 \leq \mathbb{E}\Vert \Bar{\boldsymbol\omega}_{r} - \boldsymbol\omega _{\ast}^{\varrho} \Vert^2,
\label{eqTheorem2}
\end{equation}

where $\boldsymbol\omega _{\ast}^{\varrho}$ represents the optimized weight with $p^\varrho$ distribution to fit the dataset. $\boldsymbol\omega_{r}^\varrho$ is the received local model with $p^\varrho$ distribution. And finally, $\bar{\boldsymbol\omega}_{r}$ defines the {$\textsc{FedAvg}$} uniform global model in round $r$.

{\bf{\textsc{Proof.}}} Using induction, the result is proven. First, we include the following two relationships:
\begin{equation}
\bar{\boldsymbol\omega}_{_{r=1}}=\boldsymbol\omega_{0}-\eta \nabla \bar{SG}_{r=1},
\label{eqProo2}
\end{equation}
\begin{equation}
{\boldsymbol\omega}_{_{r=1}}^\varrho=\boldsymbol\omega_{0}-\eta \nabla {L-BFGS}_{r=1}^\varrho,
\label{eqProo3}
\end{equation}

On the server side and after the first round, {$\textsc{FedAvg}$} gradients are displayed with $\bar{SG}_{r=1}$ and {$\textsc{FedAnil}$} gradients with ${L-BFGS}_{r=1}^\varrho$. The execution of the first round of {$\textsc{SGD}$} with {$\textsc{FedAvg}$} is given in \eqref{eqProo2} and the execution of the first round of {$\textsc{L-BFGS}$} with {$\textsc{FedAnil}$} is given in \eqref{eqProo3}. Therefore, according to \eqref{eqProo2} and \eqref{eqProo3}, we can conclude \eqref{eqProo4}:
\begin{equation}
\mathbb{E}\Vert \boldsymbol\omega_{r=1}^\varrho - \boldsymbol\omega _{\ast}^{\varrho} \Vert^2 \leq \mathbb{E}\Vert  \bar{\boldsymbol\omega}_{r=1} - \boldsymbol\omega _{\ast}^{\varrho} \Vert^2.
\label{eqProo4}
\end{equation}

After this, it is assumed that\eqref{eqProo5} is correct in $r^{th}$ round, we will have:
\begin{equation}
\mathbb{E}\Vert \boldsymbol\omega_{r}^\varrho - \boldsymbol\omega _{\ast}^{\varrho} \Vert^2 \leq \mathbb{E}\Vert  \bar{\boldsymbol\omega}_{r} - \boldsymbol\omega _{\ast}^{\varrho} \Vert^2.
\label{eqProo5}
\end{equation}

Now, using the \eqref{eqProo2} and \eqref{eqProo3}, we check the round $(r+1)^{th}$ and then the \eqref{eqProo6} is obtained:
\begin{equation}
\mathbb{E}\Vert \boldsymbol\omega_{r}^\varrho - \eta \nabla \bar{SG}_{r} - \boldsymbol\omega _{\ast}^{\varrho} \Vert^2 \leq \mathbb{E}\Vert  \bar{\boldsymbol\omega}_{t} - \eta \nabla {L-BFGS}_{r}^\varrho - \boldsymbol\omega _{\ast}^{\varrho} \Vert^2,
\label{eqProo6}
\end{equation}

And finally, the \eqref{eqProo7} can be expressed:
\begin{equation}
	\mathbb{E}\Vert \boldsymbol\omega_{r+1}^\varrho - \boldsymbol\omega _{\ast}^{\varrho} \Vert^2 \leq \mathbb{E}\Vert  \bar{\boldsymbol\omega}_{t+1} - \boldsymbol\omega _{\ast}^{\varrho} \Vert^2.
	\label{eqProo7}
\end{equation}

Convergence in training local models means that the optimized parameters correctly match the training dataset pattern of local enterprises. The trained model will be approved if the value of the parameters shows a stable trend or is within an acceptable error range. It is possible to theoretically show the logic of training local models if we prove that in \emph{L-BFGS} or {$\textsc{Adam}$}, the selection of the weight parameter converges to an acceptable range. However, because the value of the weight parameter is chosen randomly in the initial stages of model training and the optimal weight is not known, convergence is measured using {\textsc{Gradient Descent}}. If the $F(\boldsymbol\omega)$ continuously converges to $0$, the trained model converges to the optimal model. Therefore, the optimal and final goal of convergence in non-IID data is to converge each non-IID data to the optimal model individually.

\section{Experiments}\label{EP1}
This section will compare and evaluate the {$\textsc{FedAnil}$} model with seven related and well-known baseline models, {\textsc{ShieldFL}} \citep{ref2}, {\textsc{RVPFL}} \citep{ref19}, {\textsc{RFA}} \citep{ref20}, {\textsc{FedAdam}} \citep{ref16}, {\textsc{FedProx}} \citep{ref10}, and {\textsc{FedAvg}} \citep{ref14} under different scenarios.

\subsection{Experimental Setup}
\begin{figure*}[!t]
	\centering
	\subfloat[]{\includegraphics[width=2.0in]{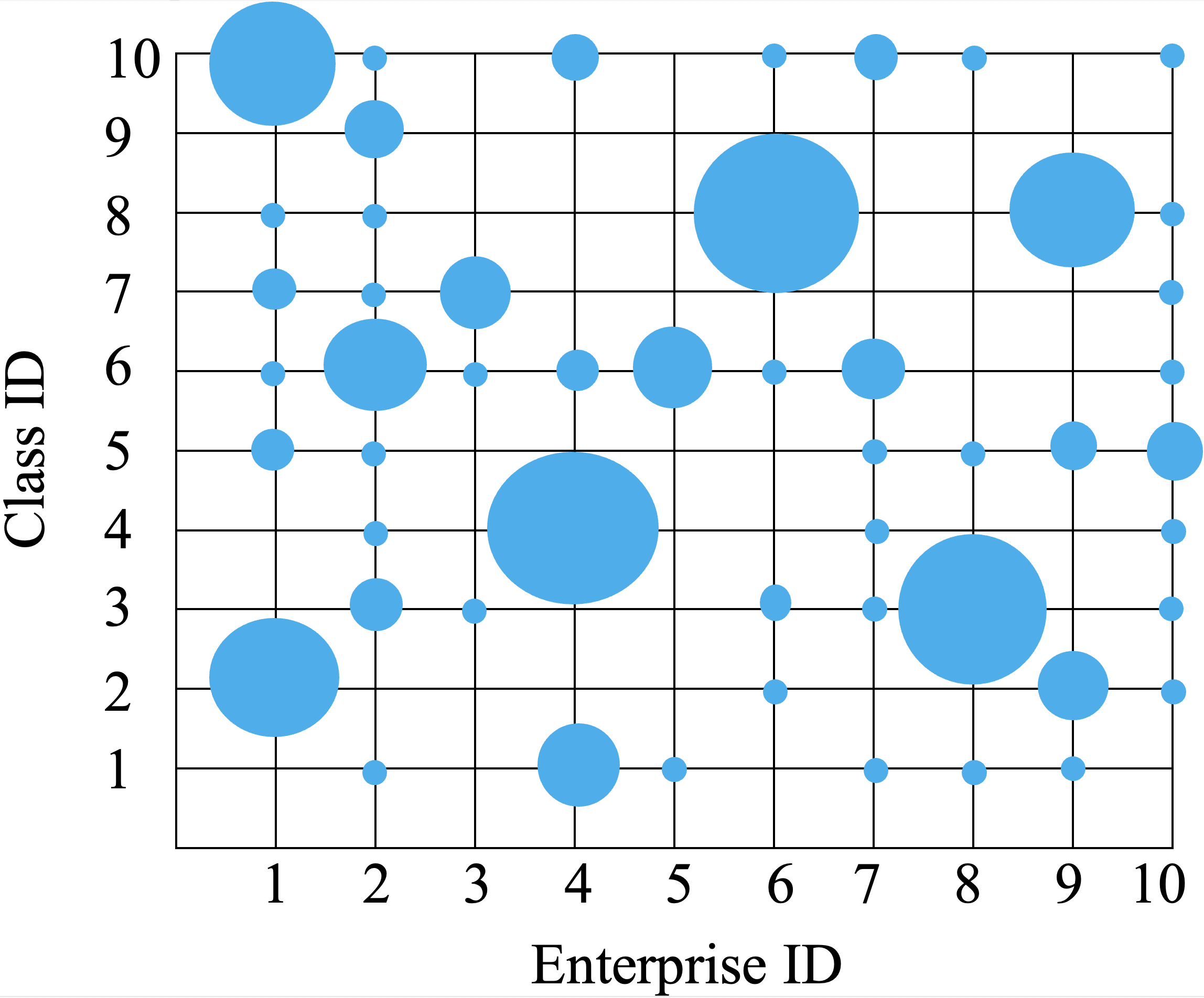}%
		\label{fig_Dricklet1}}
	\hfil
	\subfloat[]{\includegraphics[width=2.0in]{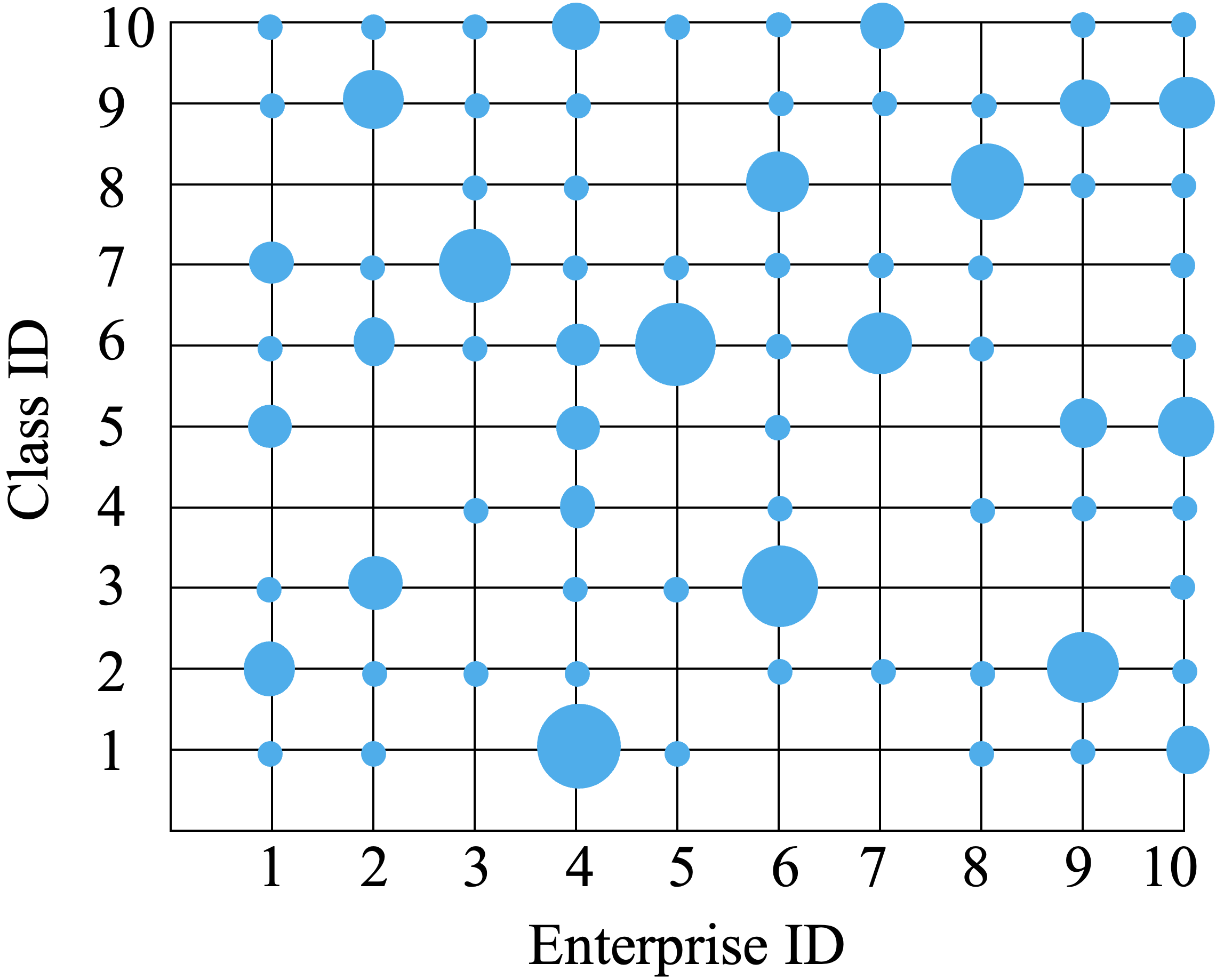}%
		\label{fig_Dricklet2}}
	\hfil
	\subfloat[]{\includegraphics[width=2.0in]{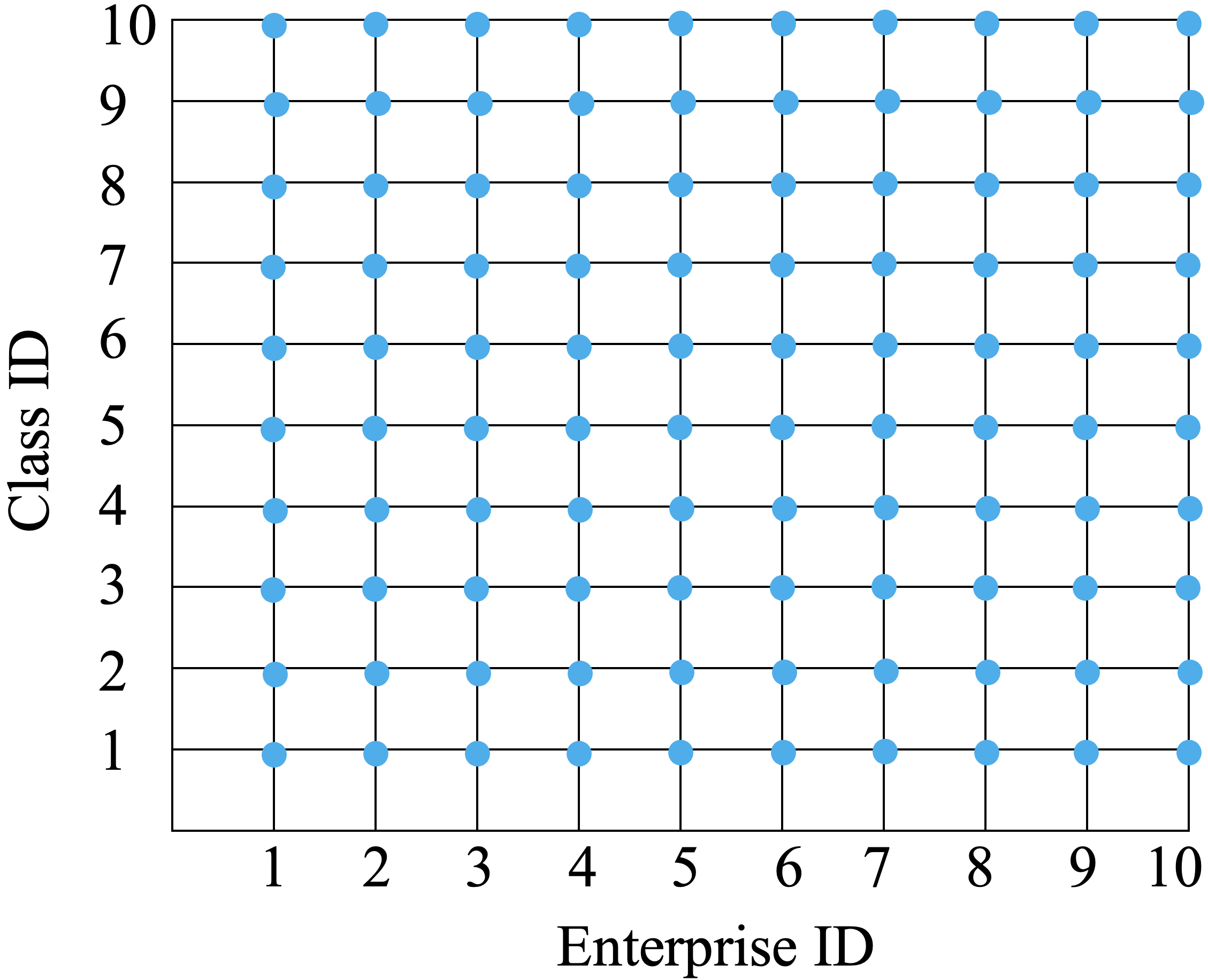}%
		\label{fig_Dricklet3}}
	\hfil
	\caption{Distribution of non-IID data with different $\alpha$, which shows the number of samples in each class for each enterprise. (a) $\alpha \rightarrow 0.1$; Non-IID distribution with extreme degree. (b) $\alpha \rightarrow 1$; Non-IID distribution with moderate degree. (c) $\alpha \rightarrow\infty$; Data distribution with very weak degree (IID).}
	\label{fig_Dricklet}
\end{figure*}

There is a multi-dimensional continuous distribution in probability statistics called Dirichlet distribution \citep{ref7}, which is known as Dir ($\alpha$), and $p \thicksim Dir (\alpha), \alpha > 0$. This distribution is used to generate non-IID datasets. To control the level of each enterprise's non-IID data distribution, parameter $\alpha$ is used. As shown in Fig. \ref{fig_Dricklet1}, the $\alpha$ lower, the higher the degree of skewness and non-IID distribution. Conversely, the larger the parameter value $\alpha$ is, the lower the degree of skew and similar to IID (Fig. \ref{fig_Dricklet3}). In the $\textsc{FedAnil}$, we partition the entire dataset using the Dirichlet distribution ($\alpha=0.1$). In this way, each local enterprise has different class types, and the total samples in each local dataset will also differ. Therefore, the $\textsc{FedAnil}$ model considers two different modes for non-IID data:

\emph{Label distribution skew}: Local enterprises have different samples, but the number of samples is the same \citep{ref15}.

\emph{Feature distribution skew}: Local enterprises have the same samples but different features \citep{ref15}.

Since the non-IID data in the $\textsc{FedAnil}$ model is divided into two classes, benign and attack, the number of classes in the CIFAR-10 dataset is $10$. Therefore, according to the $\alpha$ parameter, in a benign class, the number of benign classes is selected. In an attack class, the attack class is randomly determined from among the remaining classes.

{\bf{\textsc{Implementation.}}} We performed our experiments using Python $3.11$ on macOS Ventura $13.2$ (22D49). PyTorch \citep{ref25} has been used to train local models of enterprises in $\textsc{FedAnil}$. Performance evaluation of the $\textsc{FedAnil}$ model has been done on the OARF benchmark \citep{ref26}. According to \citep{ref14}, the experimental setup of FDL for our $\textsc{FedAnil}$ model is shown in Table \ref{tab3}.

\begin{table}
\normalsize
	\caption{The default settings for $\textsc{FedAnil}$ model training.}
	\setlength{\tabcolsep}{1.4\tabcolsep}
	\centering
  \begin{tabular}{ p{4.0cm}|p{6.5cm}}
		\toprule
		\textbf{Hyperparameters} & \textbf{Description} \\
		\midrule
		\emph{Adam$\bigl(\beta_{_1},\beta_{_2}=0.9\bigl)$} & ResNet50 optimizer for the loss function\\
		\emph{L-BFGS} & CNN optimizer for the loss function\\
		$R=50$ & Maximum communication rounds\\
		$B_1=64$ & Local training batch size\\
		$B_2=128$ & Local testing batch size\\
		$\varepsilon=30$ & Number of \emph{Epochs}\\
  		$\mu=20\%$ & Malicious rate\\
		$\eta=0.01$, $C=100$ & Learning rate, Enterprises\\
		$\alpha=0.1$ & Non-IID data distribution rate\\
		$\varphi=0.2$ & WGAN loss function threshold\\
		\bottomrule
	\end{tabular}
	\label{tab3}
\end{table}

{\bf{\textsc{Dataset and Models.}}} In Table \ref{tab4} the models and dataset used to evaluate the $\textsc{FedAnil}$ by OARF benchmark are shown. The $\textsc{FedAnil}$ will be evaluated under diverse models and non-IID datasets according to Table \ref{tab4}.

\begin{enumerate}
	\item \emph{Sentiment analysis.} Sent140 dataset \citep{ref28} was used for sentiment analysis. A linear model using average $\textsc{GloVe embeddings}$ \citep{ref29} of tweet words was used. Also, the binary logistic loss has been used to train the model according to the OARF benchmark.
 	\item \emph{Image classification.} The Fashion-MNIST dataset \citep{ref65}was used for image classification. This dataset contains 70,000 low-resolution grayscale images ($28*28$ pixels) from 10 categories, including T-shirts, dresses, bags, etc. The fashion-MNIST dataset has 10K test images, $60$K training images, and $10$ classes. CNN has also been used to train the model.
 	\item \emph{Handwritten character recognition.} The FEMNIST dataset \citep{ref27} has been used to analyze handwritten characters. In this dataset, each image is designed as a $28*28$ grayscale. Two ConvNet (CNN) models and a linear model $\phi\bigl(m;w\bigl)=w^Tm$ have been used. The CNN model is designed with two fully connected (FC) and two convolutional (Conv) layers. Also, the loss function defined by the OARF benchmark is $f\bigl(w;\bigl(m,n\bigl)\bigl)=\Gamma\bigl(n,\phi\bigl(m;w\bigl)\bigl)$. In this analysis, $\Gamma$ is defined as Multinomial Logistic Regression (MLR).
	\item \emph{Image classification.} The CIFAR-10 dataset \citep{ref30} was used for image classification. In this dataset, each image is designed as $32*32$ on a color scale. This dataset has 10K test images, $50$K training images, and $10$ classes. ResNet-50 has also been used to train the model according to the OARF benchmark. Therefore, five layers of convolutional and three layers of fully connected are used in each ResNet-50 model.
\end{enumerate}

\begin{table}
\normalsize
	\caption{The dataset used in the $\textsc{FedAnil}$}.
	\setlength{\tabcolsep}{1.7\tabcolsep}
	\centering
	\begin{tabular}{p{2.5cm}|p{1.7cm}|p{3.0cm}|p{1.0cm}|p{1.0cm}|p{1.0cm}}
		\toprule
		\textbf{dataset} & \textbf{Model} & \textbf{Task} & \textbf{\#Class}& \textbf{\#Train}& \textbf{\#Test}\\
		\midrule
  		Sent140 & GloVe & Sentiment Analysis & $2$ & $57K$ & $15K$\\
            Fashion-MNIST & CNN & Image Classification & $10$ & $60K$ & $10K$\\
		FEMNIST & CNN & Character-level & $62$ & $49K$ & $4.9K$\\
		CIFAR-10 & ResNet50 & Image Classification & $10$ & $50K$ & $10K$\\
		\bottomrule
	\end{tabular}
	\label{tab4}
\end{table}

{\bf{\textsc{Evaluation Metrics.}}} Metrics definition details are given below:

\begin{enumerate}
	\item \emph{Accuracy (ACC).} The quality of trained models in $\textsc{FedAnil}$ is evaluated by this criterion. According to \eqref{eq6}, it is the ratio of correctly predicted samples to the total samples in the validation dataset. The appropriate measurement unit for calculating accuracy will be (\emph{\%}).
	\begin{equation}
		\emph{ACC} = \frac {\kappa}{{\iota}}*100.
		\label{eq6}
	\end{equation}
	
	\item \emph{Computation overhead.} To calculate the complexity of the $\textsc{FedAnil}$ model, it is considered separately on both sides of enterprises and servers. On the enterprise’s side, operations that create computation overhead include \emph{local model training}, \emph{CKKS-FHE}, and \emph{blockchain} techniques. On the other hand, there are \emph{CS}, \emph{AP}, and \emph{WGAN} operations on the server side. The highest computation overhead on the client side equals $O(N^2)$ for \emph{CKKS-FHE}. On the server side, the highest computation overhead is related to the \emph{CS} and \emph{AP} operations, which, because they are performed serially and one after the other, is $O(N^2)$. Finally, the complexity of the $\textsc{FedAnil}$ model is $O(N^2)$. The appropriate measurement unit for calculating computation overhead will be minute (\emph{min}). This metric is calculated by \eqref{eq5-4}, \eqref{eq5-5}, and \eqref{eq5-6}.
\end{enumerate}
\begin{equation}
	\begin{split}
		\emph{COMP}_{_{C\_Side}}=\sum _{r=1}^{R}\sum _{k=1}^{\Delta c} \Bigl({T}_{{LT_r}} + {T}_{_{CKKS_r}}\Bigl).
	\end{split}
	\label{eq5-4}
\end{equation}
\begin{equation}
	\begin{split}
		\emph{COMP}_{_{S\_Side}}=\sum _{r=1}^{R} \Bigl({T}_{_{CS_{r}}} + {T}_{_{WGAN_{r}}} + {T}_{_{Agg_{r}}}\Bigl).
	\end{split}
	\label{eq5-5}
\end{equation}
\begin{equation}
	\begin{split}
		\emph{COMP}_{_{Total}}= 
		\Bigl(\emph{COMP}_{_{C\_Side}} + \emph{COMP}_{_{S\_Side}}\Bigl).
	\end{split}
	\label{eq5-6}
\end{equation}
In \eqref{eq5-4}, \eqref{eq5-5}, and \eqref{eq5-6}:
\begin{itemize}
	\item \emph{${{COMP}_{_{C\_Side}}}$}: Local enterprise's computation cost.
	\item \emph{${T}_{{LT_r}}$}: Local training computation time in round \emph{r}.
	\item \emph{{${T}_{_{CKKS_r}}$}}: \emph{CKKS-FHE} computation time. 
	\item \emph{${{COMP}_{_{S\_Side}}}$}: The computation costs on the server side. 
	\item \emph{${T}_{_{CS_{r}}}$}: \emph{CS} computation time. 
	\item \emph{${T}_{_{WGAN_{r}}}$}, {${T}_{_{Agg_{r}}}$}: \emph{WGAN} and \emph{Agg} computation time. 
	\item \emph{${{COMP}_{_{Total}}}$}: Total computing cost. 
\end{itemize}

\subsection{Attack Scenario}
In this section, {$\textsc{general}$} scenarios and {$\textsc{concrete}$} scenarios for poisoning and inference attacks are described.

{\bf{\textsc{General and Concrete scenarios.}}} The general scenario means the same threat model/attack model where the training data privacy is violated against poisoning and inference attacks. $\textsc{general scenarios}$ are shown in Table \ref{tab6}to Table \ref{tab8}.Also, a $\textsc{concrete scenario}$, which is an example scenario, means that if the enterprise is attacked, the resistance and efficiency of the $\textsc{FedAnil}$ model should be by the concrete scenario as shown in Table \ref{tab9}.

\begin{table}
\normalsize
	\caption{General scenario for the data poisoning attack.}
	\setlength{\tabcolsep}{1.7\tabcolsep}
	\centering
   \begin{tabular}{ p{2.49cm}|p{3.8cm}|p{6.5cm}}
		\toprule
		\textbf{Access point} & \textbf{Agent} & \textbf{Threat model}\\
		\midrule
		Local Enterprises & An intruder that has root access to the clients. & Malicious enterprises can send poisoning gradients to the server, affecting the global model accuracy.\\
		\bottomrule
	\end{tabular}
	\label{tab6}
\end{table}

\begin{table}
\normalsize
	\caption{General scenario for the collude attack.}
	\setlength{\tabcolsep}{1.7\tabcolsep}
	\centering
   \begin{tabular}{ p{2.49cm}|p{3.8cm}|p{6.5cm}}
		\toprule
		\textbf{Access point} & \textbf{Agent} & \textbf{Threat model}\\
		\midrule
		Local Enterprises & An intruder that intends to carry out collude attacks with the cooperation of other intruders. & Malicious enterprises that cooperate by entering poisoning gradients intend to carry out a collude attack to affect the global model accuracy.\\
		\bottomrule
	\end{tabular}
	\label{tab7}
\end{table}

\begin{table}
\normalsize
	\caption{General scenario for the model poisoning, membership inference, and reconstruction attacks.}
	\setlength{\tabcolsep}{1.7\tabcolsep}
	\centering
   \begin{tabular}{ p{2.49cm}|p{3.8cm}|p{6.5cm}}
		\toprule
		\textbf{Access point} & \textbf{Agent} & \textbf{Threat model}\\
		\midrule
		Server & An intruder that has root access to the server. & An honest but curious server cannot manipulate local parameters but can inspect all of them.\\
		\bottomrule
	\end{tabular}
	\label{tab8}
\end{table}

\begin{table}
	\caption{Concrete scenario for the Best (B), Average (A), and Worst case (W). (States=S, Collude Attack=CA, Data Poisoning=DP, Model Poisoning=MP, Membership Inference=MI, and Reconstruction Attack=RA, Accuracy=ACC.}
	\begin{tabular}{p{0.5cm}p{1.0cm}p{1.0cm}p{1.0cm}p{1.0cm}p{1.0cm}p{2.1cm}p{2.1cm}p{2.1cm}p{1.7cm}}
		\toprule
		\multirow{2}{*}{S} &
		\multicolumn{5}{c}{Attacks} &
		\multicolumn{2}{c}{Metrics} &
		\multicolumn{1}{c}{State} \\
		& {DP} & {CA} & {MP} & {MI} & {RA} & {ACC(\bf\%)} & {TcP(\bf{min})} & {B/A/W} \\
		\midrule
		1 &  \textcolor{ForestGreen}{\cmark} & \textcolor{red}{\xmark} & \textcolor{red}{\xmark} & \textcolor{red}{\xmark} & \textcolor{red}{\xmark} & $\geq77$ & $\leq160$ & B\\
		2 & \textcolor{red}{\xmark} &  \textcolor{ForestGreen}{\cmark} & \textcolor{red}{\xmark} & \textcolor{red}{\xmark} & \textcolor{red}{\xmark} & $\geq79$ & $\leq150$ & B\\
		3 & \textcolor{red}{\xmark} & \textcolor{red}{\xmark} &  \textcolor{ForestGreen}{\cmark} & \textcolor{red}{\xmark} & \textcolor{red}{\xmark} & $\geq75$ & $\leq170$ & B\\
		4 & \textcolor{red}{\xmark} & \textcolor{red}{\xmark} & \textcolor{red}{\xmark} &  \textcolor{ForestGreen}{\cmark} & \textcolor{red}{\xmark} & $\geq80$ & $\leq130$ & B\\
		5 & \textcolor{red}{\xmark} & \textcolor{red}{\xmark} & \textcolor{red}{\xmark} & \textcolor{red}{\xmark} &  \textcolor{ForestGreen}{\cmark} & $\geq82$ & $\leq120$ & B\\
		6 &  \textcolor{ForestGreen}{\cmark} &  \textcolor{ForestGreen}{\cmark} &  \textcolor{ForestGreen}{\cmark} & \textcolor{red}{\xmark} & \textcolor{red}{\xmark} & $\geq65$ & $\leq240$ & A\\
		7 & \textcolor{red}{\xmark} & \textcolor{red}{\xmark} & \textcolor{red}{\xmark} &  \textcolor{ForestGreen}{\cmark} &  \textcolor{ForestGreen}{\cmark} & $\geq70$ & $\leq200$ & A\\
		8 &  \textcolor{ForestGreen}{\cmark} &  \textcolor{ForestGreen}{\cmark} &  \textcolor{ForestGreen}{\cmark} &  \textcolor{ForestGreen}{\cmark} &  \textcolor{ForestGreen}{\cmark} & $\geq60$ & $\leq260$ & W\\

    \bottomrule
  \end{tabular}
    \label{tab9}
\end{table}

\subsection{Experimental Results}
\begin{figure*}[!t]
	\centering
	\subfloat[]{\includegraphics[width=1.5in]{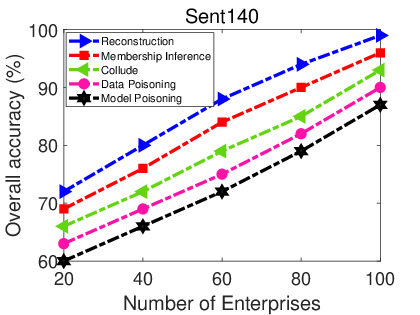}%
		\label{fig_accuracy_1_case}}
	\hfil
 \subfloat[]{\includegraphics[width=1.5in]{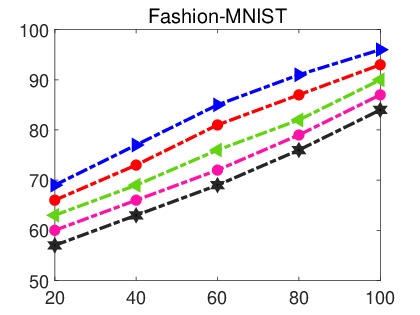}%
		\label{fig_accuracy_19_case}}
	\hfil
	\subfloat[]{\includegraphics[width=1.5in]{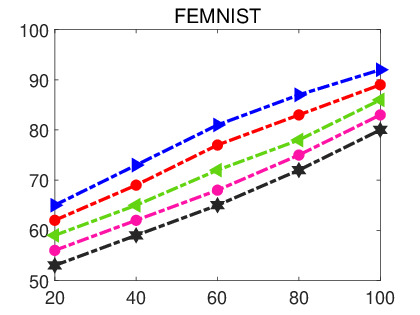}%
		\label{fig_accuracy_5_case}}
	\hfil
	\subfloat[]{\includegraphics[width=1.5in]{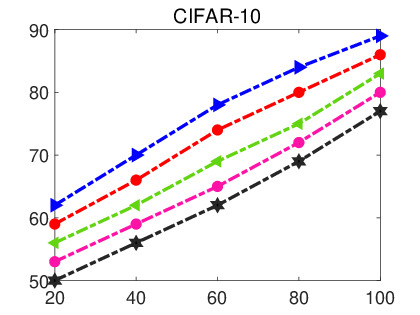}%
		\label{fig_accuracy_9_case}}
	\hfil
	\subfloat[]{\includegraphics[width=1.5in]{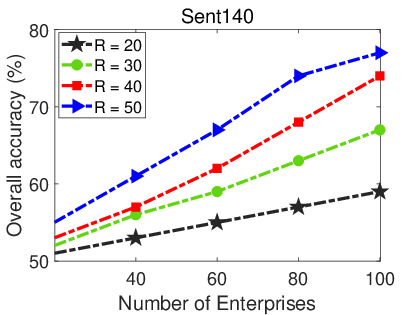}%
		\label{fig_accuracy_2_case}}
	\hfil
 \subfloat[]{\includegraphics[width=1.5in]{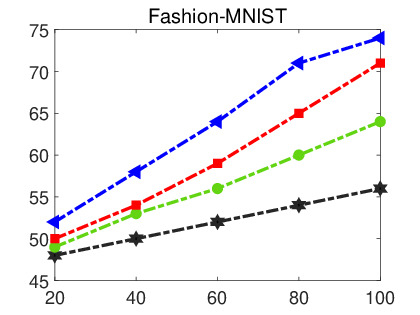}%
		\label{fig_accuracy_100_case}}
	\hfil
	\subfloat[]{\includegraphics[width=1.5in]{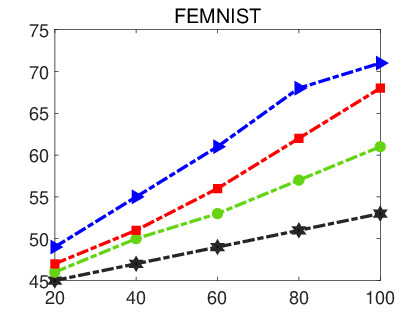}%
		\label{fig_accuracy_6_case}}
	\hfil
	\subfloat[]{\includegraphics[width=1.5in]{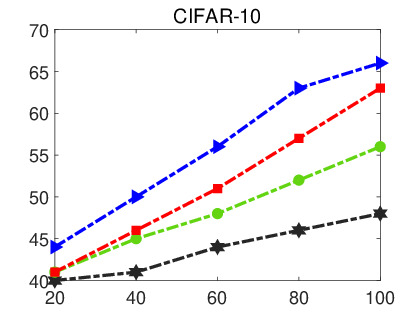}%
		\label{fig_accuracy_10_case}}
	\hfil
	\subfloat[]{\includegraphics[width=1.5in]{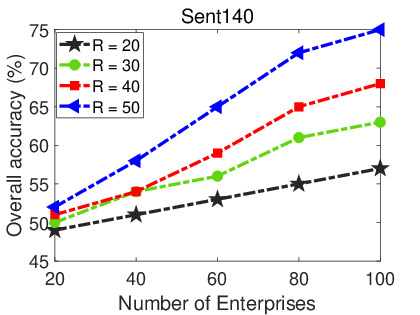}%
		\label{fig_accuracy_3_case}}
	\hfil
 \subfloat[]{\includegraphics[width=1.5in]{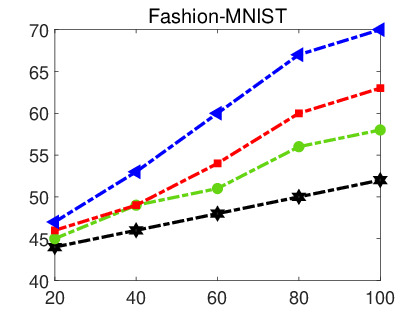}%
		\label{fig_accuracy_111_case}}
	\hfil
	\subfloat[]{\includegraphics[width=1.5in]{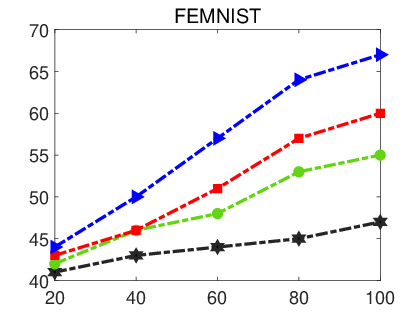}%
		\label{fig_accuracy_7_case}}
	\hfil
	\subfloat[]{\includegraphics[width=1.5in]{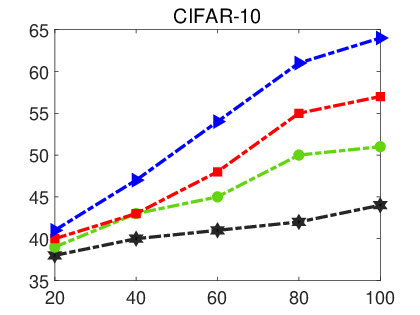}%
		\label{fig_accuracy_11_case}}
	\hfil
	\subfloat[]{\includegraphics[width=1.5in]{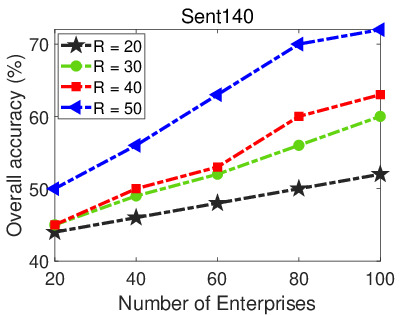}%
		\label{fig_accuracy_4_case}}
	\hfil
 \subfloat[]{\includegraphics[width=1.5in]{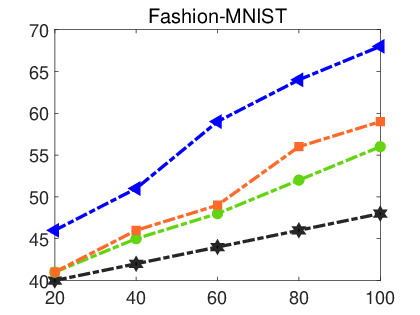}%
		\label{fig_accuracy_112_case}}
	\subfloat[]{\includegraphics[width=1.5in]{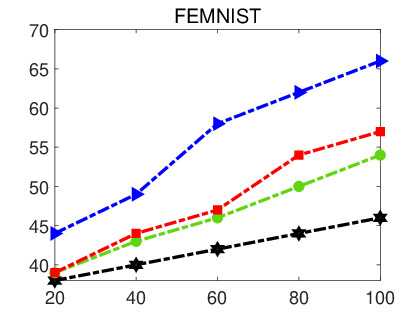}%
		\label{fig_accuracy_8_case}}
	\hfil
	\subfloat[]{\includegraphics[width=1.5in]{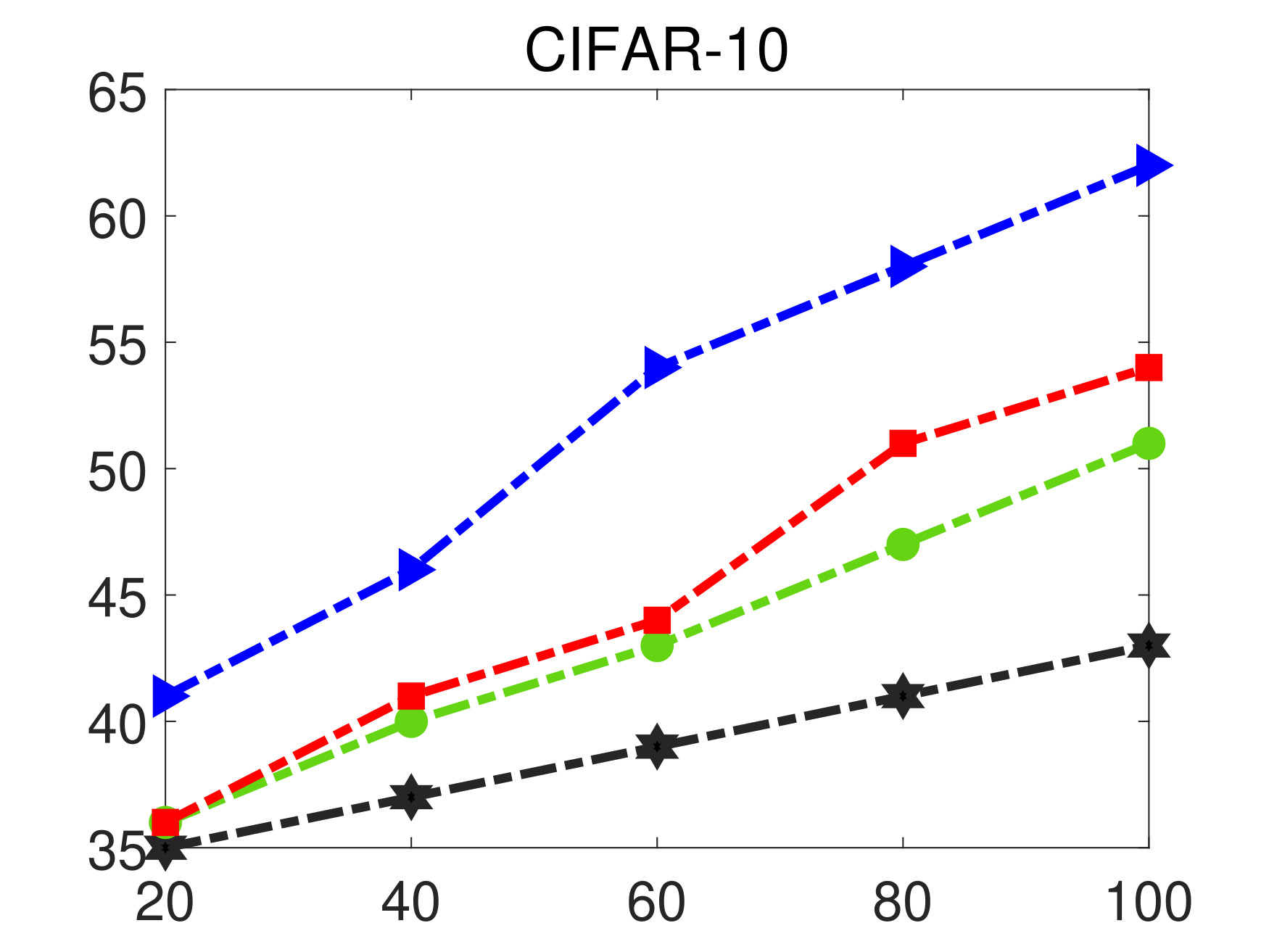}%
		\label{fig_accuracy_12_case}}
        \hfil
	\caption{Comparison of the overall accuracy in $\textsc{FedAnil}$, where the $\varepsilon$ is fixed at $30$ and $\mu$=20\%. Best case ((a, b, c, and d) for $R=50$). Average case (e, f, g, and h): Poisoning attacks. Average case (i, j, k, and l): Inference attacks. Worst case (m, n, o, and p): Poisoning and inference attacks.}
	\label{fig_self_accuracy}
\end{figure*}

\begin{figure*}[!t]
	\centering
	\subfloat[]{\includegraphics[width=1.5in]{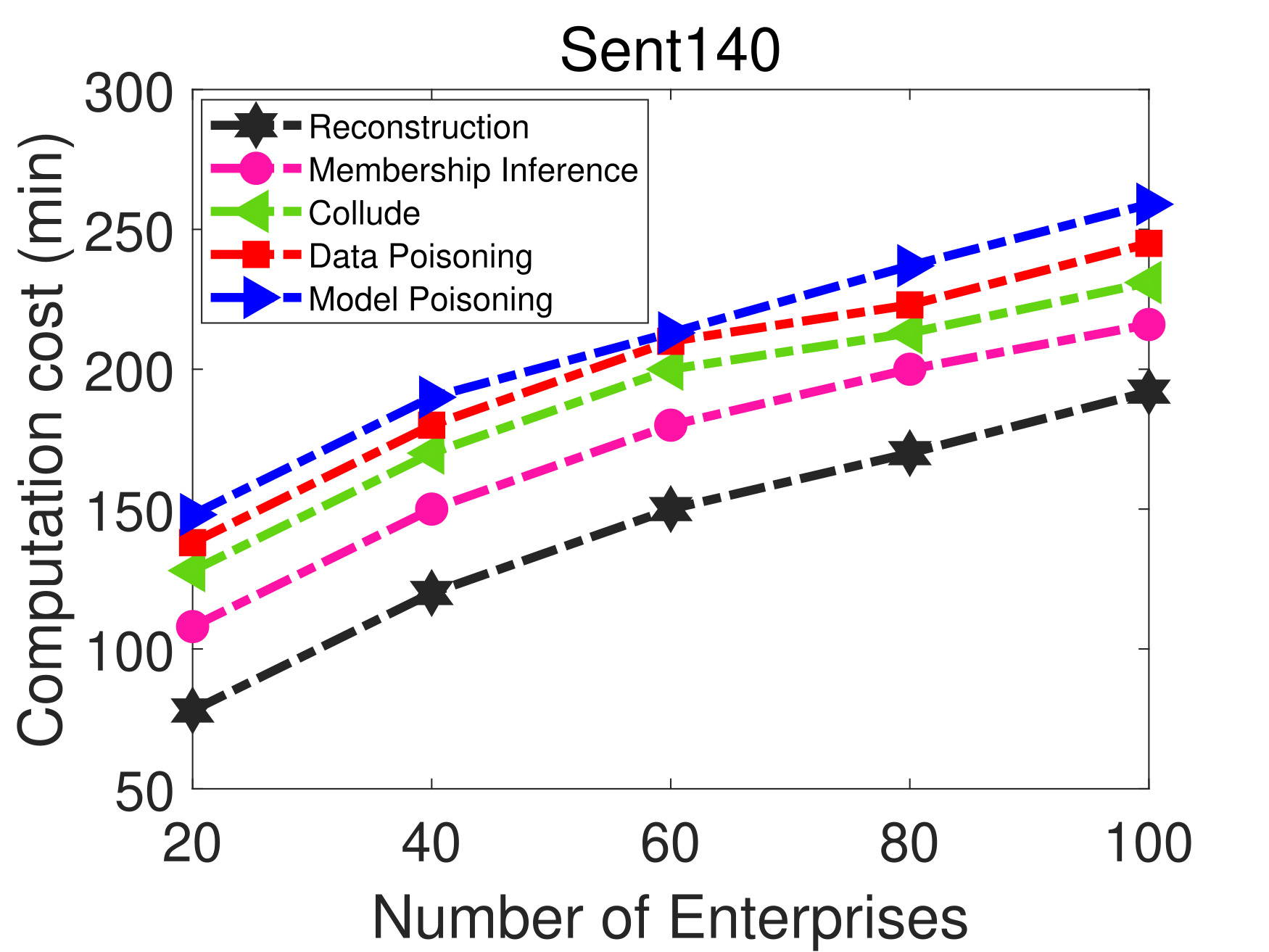}%
		\label{fig_computation_1111_case}}
	\hfil
 \subfloat[]{\includegraphics[width=1.5in]{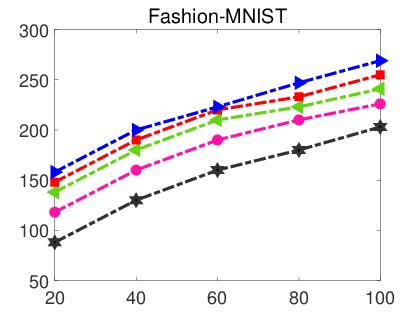}%
		\label{fig_computation_19_case}}
	\hfil
	\subfloat[]{\includegraphics[width=1.5in]{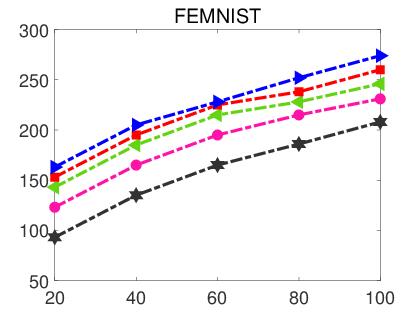}%
		\label{fig_computation_5_case}}
	\hfil
	\subfloat[]{\includegraphics[width=1.5in]{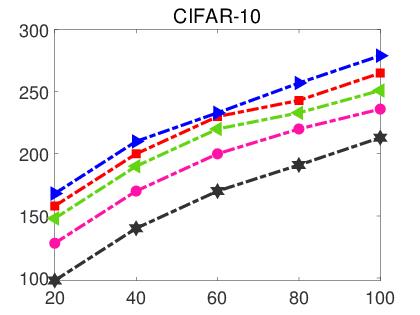}%
		\label{fig_computation_9_case}}
	\hfil
	\subfloat[]{\includegraphics[width=1.5in]{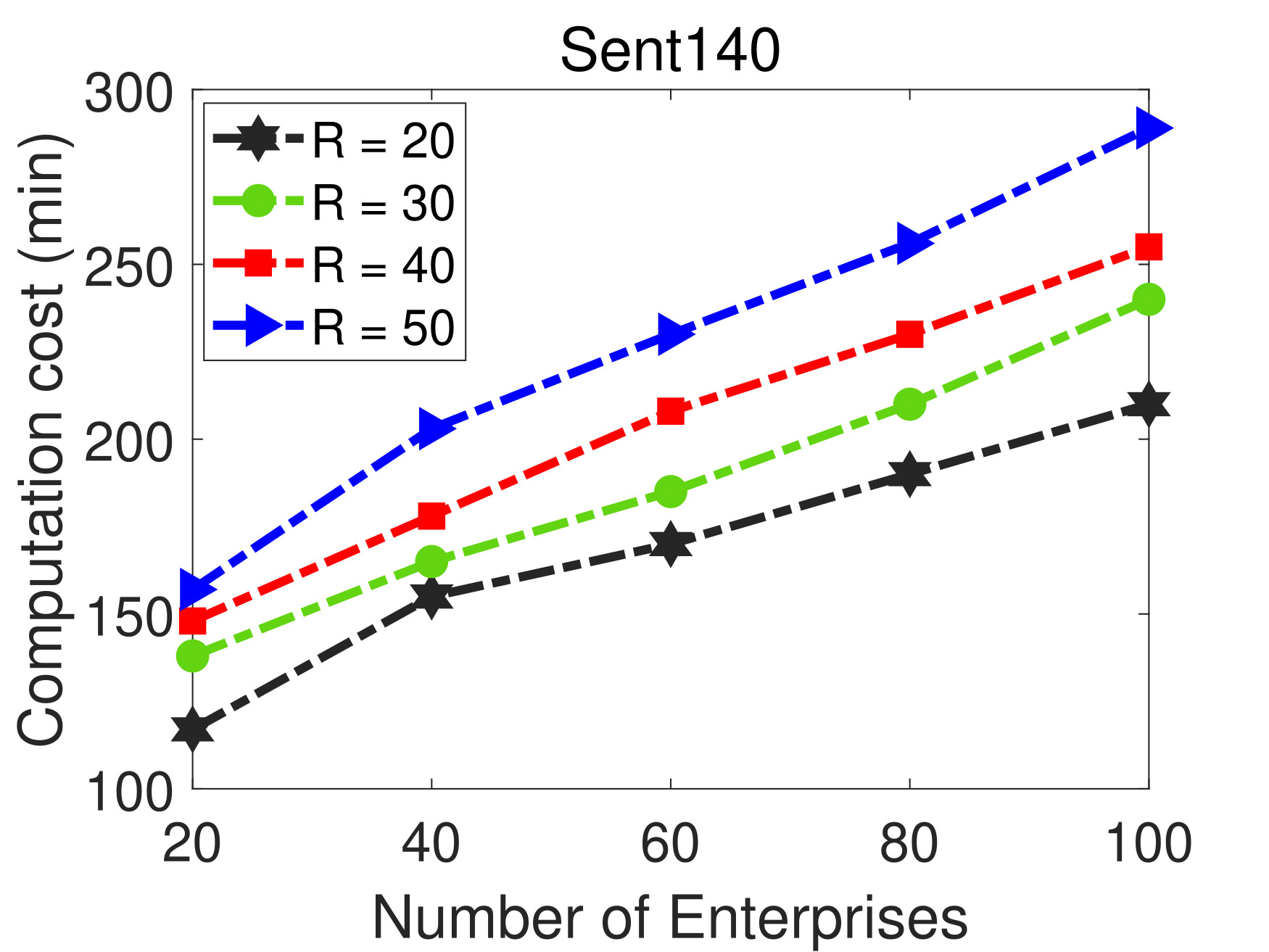}%
		\label{fig_computation_2_case}}
	\hfil
 \subfloat[]{\includegraphics[width=1.5in]{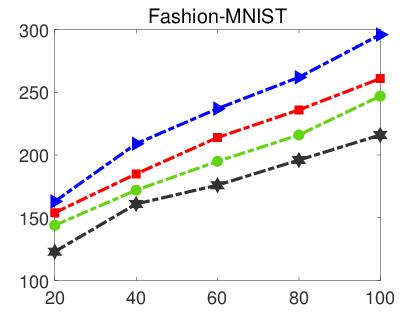}%
		\label{fig_computation_100_case}}
	\hfil
	\subfloat[]{\includegraphics[width=1.5in]{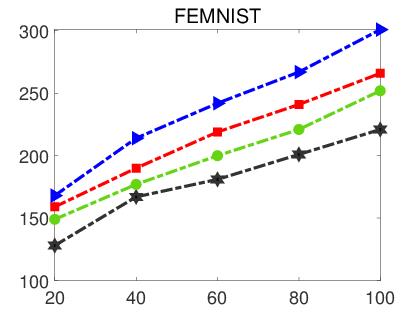}%
		\label{fig_computation_6_case}}
	\hfil
	\subfloat[]{\includegraphics[width=1.5in]{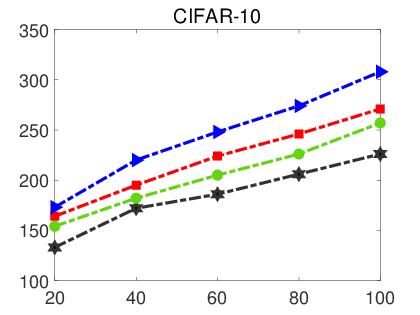}%
		\label{fig_computation_10_case}}
	\hfil
	\subfloat[]{\includegraphics[width=1.5in]{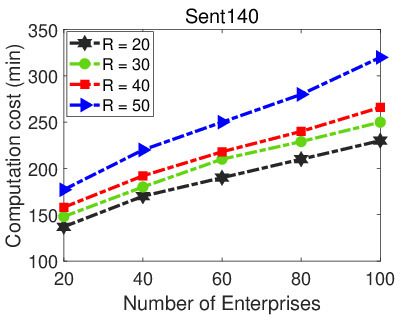}%
		\label{fig_computation_3_case}}
	\hfil
 \subfloat[]{\includegraphics[width=1.5in]{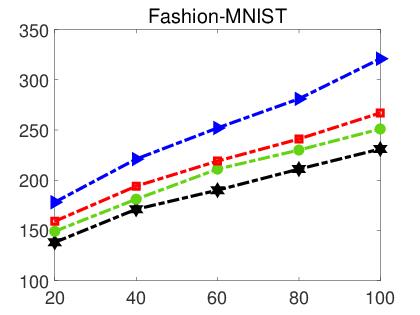}%
		\label{fig_computation_111_case}}
	\hfil
	\subfloat[]{\includegraphics[width=1.5in]{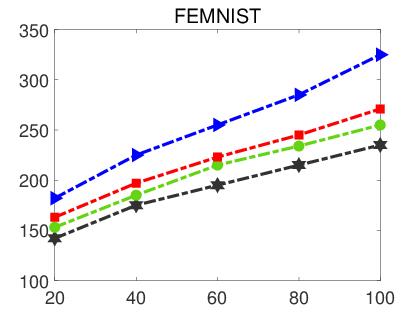}%
		\label{fig_computation_7_case}}
	\hfil
	\subfloat[]{\includegraphics[width=1.5in]{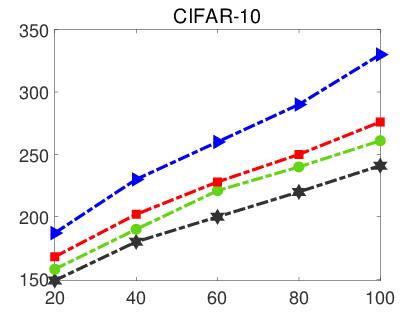}%
		\label{fig_computation_11_case}}
	\hfil
	\subfloat[]{\includegraphics[width=1.5in]{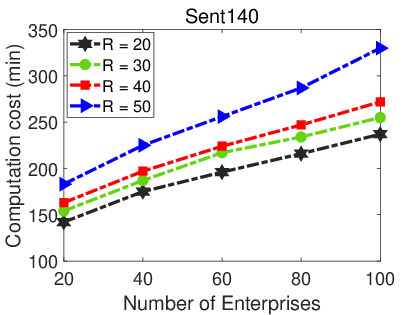}%
		\label{fig_computation_4_case}}	
 \hfil
 \subfloat[]{\includegraphics[width=1.5in]{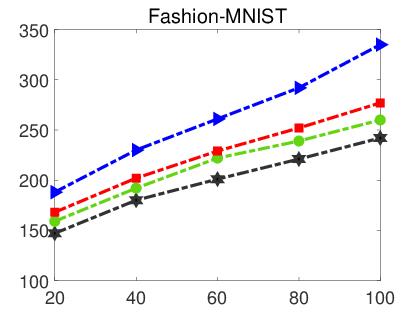}%
		\label{fig_computation_112_case}}
     \hfil
	\subfloat[]{\includegraphics[width=1.5in]{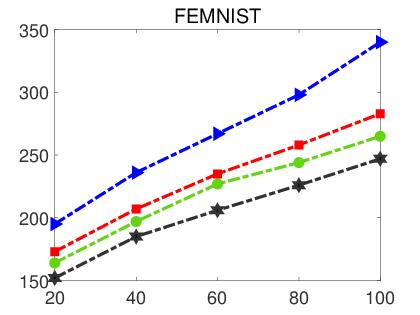}%
		\label{fig_computation_8_case}}
	\hfil
	\subfloat[]{\includegraphics[width=1.5in]{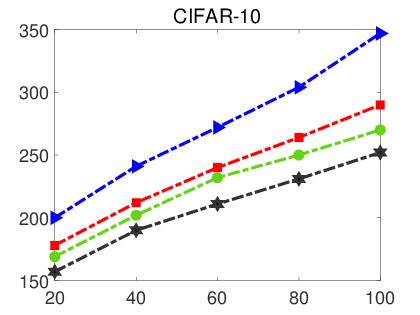}%
		\label{fig_computation_12_case}}
		\caption{Comparison of the total computation cost in $\textsc{FedAnil}$, where the $\varepsilon$ is fixed at $30$ and $\mu$=20\%. Best case ((a, b, c, and d) for $R=50$). Average case (e, f, g, and h): Poisoning attacks. Average case (i, j, k, and l): Inference attacks. Worst case (m, n, o, and p): Poisoning and inference attacks.}
	\label{fig_self_computation}
\end{figure*}

We train three popular deep-learning models on four diverse datasets, Sent140 \citep{ref28}, Fashion-MNIST \citep{ref65}, FEMNIST \citep{ref27}, and CIFAR-10 \citep{ref30}, and evaluate $\textsc{FedAnil}$ robustness based on three scenarios. The evaluation has been according to the eight states in Table \ref{tab9} (\emph{Best-case, Average-case, and Worst-case}). In the first scenario, the $\textsc{FedAnil}$ model is compared with itself in the specified metrics as denoted in Fig. \ref{fig_self_accuracy} and  Fig. \ref{fig_self_computation}. 

{\bf{\textsc{Overall Accuracy.}}} In Fig. \ref{fig_self_accuracy} (\emph{a-p}), the \emph{ACC} of the $\textsc{FedAnil}$ in four communication rounds has been compared and evaluated based on three scenarios. With the number of enterprises equal $40$ to $100$, the $\textsc{FedAnil}$ \emph{ACC} is much better. The first reason for the better \emph{ACC} of the $\textsc{FedAnil}$ model in higher rounds is that on the server side, the model of all enterprises is clustered based on data distribution by \emph{CS} and \emph{AP} techniques. Then, aggregation is done on homogeneous clusters. Therefore, the goal of the techniques based on \emph{CS} and \emph{AP} is to cluster heterogeneous models of local enterprises. This leads to the homogenization of local models, reducing the convergence time and increasing the accuracy of the global model. The second reason is that after the global model is created, it is trained by \emph{WGAN}. The \emph{WGAN} technique trains the global model on hard shadow samples ($\vartheta$). Thus, this step can solve the problem of feature distribution skew and class labels by correctly identifying the features and then correctly classifying the samples into related classes. As a result, in this case, the global model can preserve the local model's accuracy and convergence that follow their local training through this global model. The third reason for the better accuracy of the proposed model is to prevent poisoning attacks. Their goal is to increase the convergence time and ultimately reduce the accuracy of the global model. In data poisoning, the attacker manipulates the training dataset. In addition, manipulating the gradient value that the optimizer calculates in the model poisoning attack takes the model out of the optimal state so that the test error rate increases and causes a reduction in the global model \emph{ACC}.

{\bf{\textsc{Computation Overhead.}}} As shown in Fig. \ref{fig_self_computation}, $\textsc{FedAnil}$ in \emph{R = $50$} has more computational overhead than the (\emph{R = $20$}, \emph{R = $30$}, and \emph{R = $40$}). The higher computational overhead in \emph{R = $50$} compared to fewer communication rounds is that more computations are performed in each training round. In each round, \emph{Local Training} and \emph{HE} operations on the client side, as well as the computations of \emph{WGAN} and \emph{CS} clustering on the server side, are performed. Therefore, in each of these steps, computing time is used.

\begin{figure}[!t]
	\centering
	\includegraphics[width=3in]{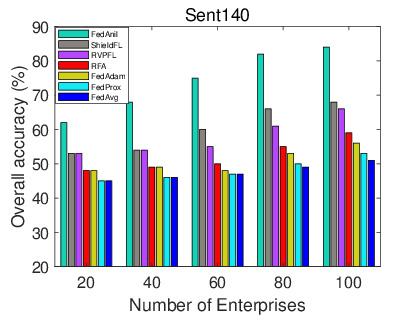}
	\caption{Comparison of the overall accuracy between $\textsc{FedAnil}$,{\textsc{ShieldFL}}, {\textsc{RVPFL}}, {\textsc{RFA}}, {\textsc{FedAdam}}, {\textsc{FedProx}}, and {\textsc{FedAvg}}, where the $\varepsilon$ is fixed at $30$ and $R=50$. (Worst case: Poisoning and inference attacks, $\mu=20\%$).}
        \label{fig_Comparission3Method1}
\end{figure}

\begin{figure}[!t]
	\centering
	\includegraphics[width=3in]{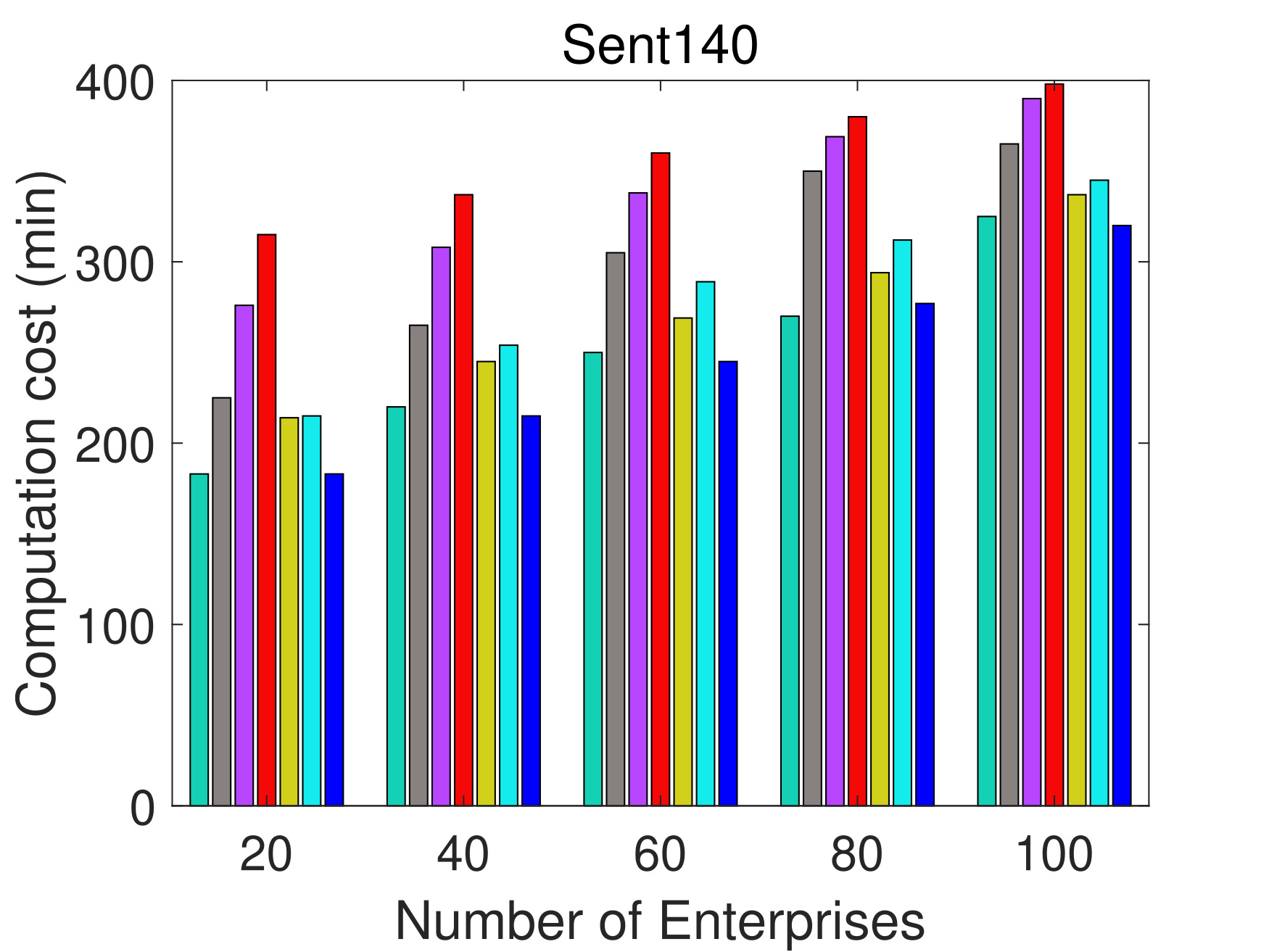}
	\caption{Comparison of the computation cost between $\textsc{FedAnil}$, {\textsc{ShieldFL}}, {\textsc{RVPFL}}, {\textsc{RFA}}, {\textsc{FedAdam}}, {\textsc{FedProx}}, and {\textsc{FedAvg}}, where the $\varepsilon$ is fixed at $30$ and $R=50$. (Worst case: Poisoning and inference attacks, $\mu=20\%$).}
	\label{fig_Comparission3Method2}
\end{figure}

{\bf{\textsc{Comparison with baseline approaches.}}} In this section, the $\textsc{FedAnil}$ model has been compared and evaluated with six approaches, {\textsc{ShieldFL}}, {\textsc{RVPFL}}, {\textsc{RFA}}, {\textsc{FedAdam}}, {\textsc{FedProx}}, and {\textsc{FedAvg}} according to Fig. \ref{fig_Comparission3Method1} and Fig. \ref{fig_Comparission3Method2}.

{\bf{\textsc{Overall Accuracy.}}} As illustrated in Fig. \ref{fig_Comparission3Method1}, with the increase in enterprises, the $\textsc{FedAnil}$ overall accuracy has increased. When the number of enterprises is from $20$ to $40$, the accuracy of the $\textsc{FedAnil}$ model does not increase uniformly and has less improvement. However, when total enterprises increase from $40$ to $100$, the overall accuracy improves. As shown in Fig. \ref{fig_self_accuracy}explained, the accuracy of the proposed model is in a better state for three reasons: first, clustering heterogeneous local models; second, training the model on hard shadow samples; third, fending off poisoning attacks to reduce the convergence time and increase the model accuracy. In {\textsc{ShieldFL}} method, model training is done on non-IID, and no approach is proposed for model training on hard (shadow) samples. Therefore, there is a possibility that the model accuracy is reduced in these samples, which is a strong reason for the poor accuracy. The logic behind the low accuracy of the {\textsc{RVPFL}} is that they used an encryption technique with much gradient loss during decryption. Also, since the efficiency of {\textsc{RFA}} on non-IID is weak, the model accuracy is low. In traditional models, the performance of {\textsc{FedAvg}}, {\textsc{FedAdam}}, and {\textsc{FedProx}} against five attacks and in non-IID data is weak and causes the model to diverge. Therefore, this has reduced the model's accuracy.

{\bf{\textsc{Computation Overhead.}}} The goal of this metric is to show the computation cost of the \emph{blockchain} \emph{HE} operations on the client side, as well as the computation of \emph{WGAN} and \emph{CS} based clustering on the server side. According to the results given in Fig. \ref{fig_Comparission3Method2}, it is that the $\textsc{FedAnil}$ model has lightweight computation due to not using heavy operations, but in contrast, {\textsc{ShieldFL}}, {\textsc{RVPFL}}, and {\textsc{RFA}} approaches used HE, adaptive CS, and lossy encryption, respectively, which have computationally heavy operations. For example, in total enterprises equal to $100$, the computation overhead of the $\textsc{FedAnil}$ model is $325$, which is $365$, $390$, and $398$ \emph{min} in the {\textsc{ShieldFL}}, {\textsc{RVPFL}}, and {\textsc{RFA}} compared approaches, respectively. On the other hand, the $\textsc{FedAnil}$ model has more computation overhead than the {\textsc{FedAvg}} method due to the use of HE operations. The reason for the low computation overhead of this method is that fewer calculations are performed on the client side, which has caused its computation overhead to be low. Although the $\textsc{FedAnil}$ model is lightweight in terms of computation overhead, it does not perform better compared to {\textsc{FedAvg}}.

\begin{figure}[!t]
	\centering
	\includegraphics[width=3in]{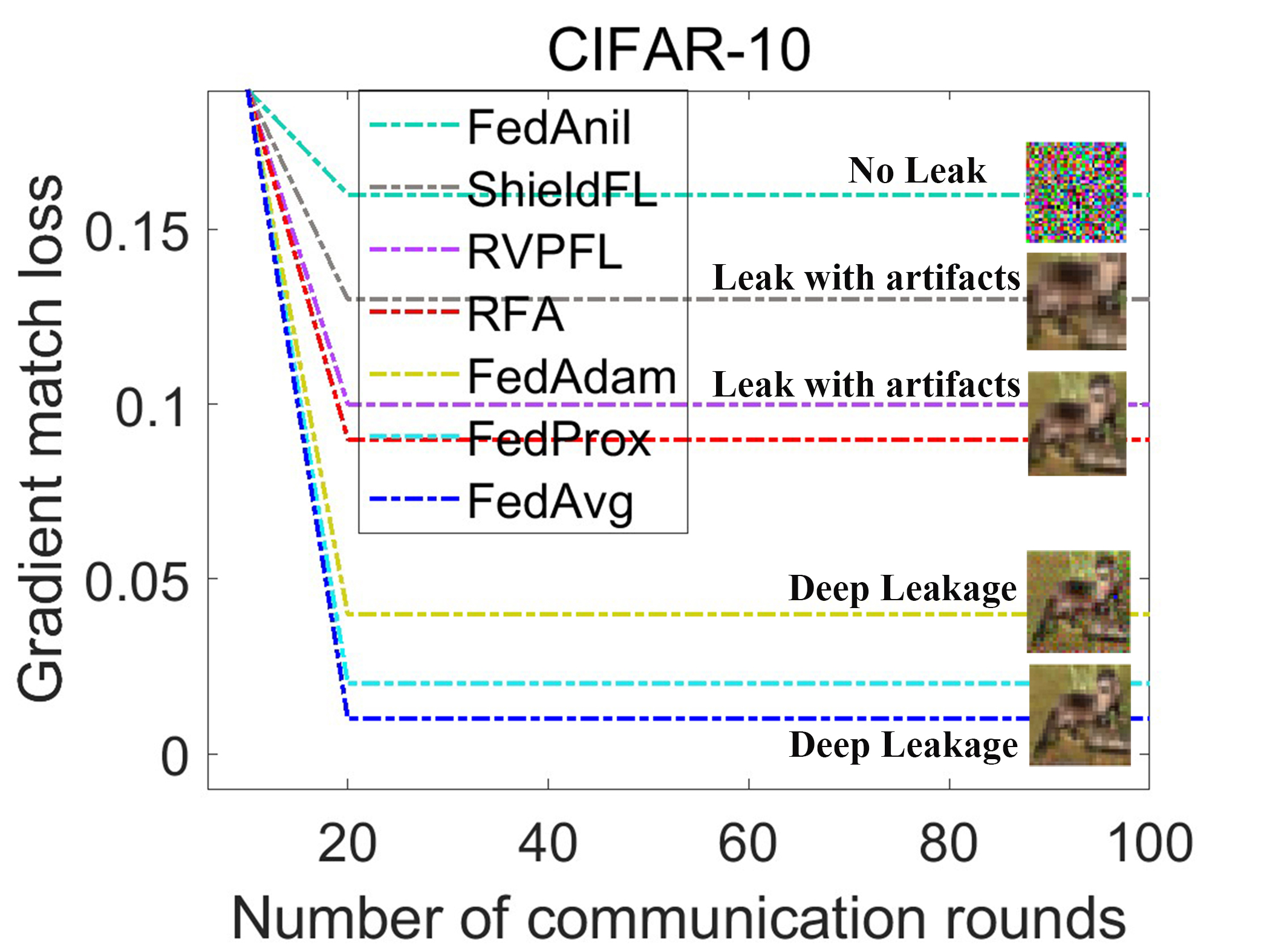}
	\caption{Comparison of the resistance to inference attacks between $\textsc{FedAnil}$, $\textsc{ShieldFL}$, $\textsc{RVPFL}$, $\textsc{RFA}$, $\textsc{FedAdam}$, $\textsc{FedProx}$, and $\textsc{FedAvg}$, where the $\varepsilon$ is fixed at $30$. (Average case: Membership inference and Reconstruction attacks, $\mu=20\%$).}
	\label{fig_Comparission3Method3}
\end{figure}

{\bf{\textsc{The effectiveness of resistance to inference attacks.}}} In $\textsc{FedAnil}$, the \emph{L-BFGS} algorithm \citep{ref70} is used as an optimizer and $\textsc{Gradient Matching Loss (GML)}$ function for all training model parameters. The \emph{GML} function represents the difference between real raw and dummy samples. Specifically, \emph{GML} means, during the optimization phase, how well the gradients updated by the benign local enterprise match the gradients fabricated by the malicious local enterprise. The intruder executes this process and will gradually match the original updated gradients. In other words, if the gradients are processed clearly and without encrypting by performing the gradient matching operation, the fake data will be close to the original data, leading to privacy leakage. The smaller the value of the \emph{GML} function, the more information is leaked (\emph{Deep Leakage}). No data is leaked when the \emph{GML} value exceeds $0.15$ (\emph{No Leak}). Fig. \ref{fig_Comparission3Method3}shows the effects of privacy leakage versus the number of communication rounds. As demonstrated in Fig. \ref{fig_Comparission3Method3}, the loss function of other approaches is less than $0.15$. Therefore, they cannot prevent the leakage of private and sensitive data of local enterprises. Moreover, the $\textsc{FedAnil}$ model preserves data privacy and does not leak any information to the intruder. The reason for the robustness of the $\textsc{FedAnil}$ model is the use of a technique based on \emph{CKKS-FHE}, where the local model parameters are exchanged encrypted and without decryption between local enterprises and the central enterprises. This makes the intruder unable to compare any output obtained from the fake data with any reference model to make privacy leak in the enterprise data by setting these parameters. Because all local model parameters are encrypted by \emph{CKKS-FHE}.

\section{Conclusions and Future Work}\label{CC1}
This paper developed a privacy-preserving FDL model to address the non-IID data (Label and Feature distribution skew) and privacy-preserving challenges. Experimental results show that, compared to the baseline approaches, the $\textsc{FedAnil}$ outperforms in terms of accuracy and computation overhead, which is a potentially effective model for privacy-preserving. These results prove that $\textsc{FedAnil}$ is resistant to poisoning and inference attacks and has better accuracy on non-IID data. The reason for the better $\textsc{FedAnil}$ accuracy is to prevent poisoning attacks, address the non-IID challenge via clustering technique, and train the global model by \emph{WGAN}. Moreover, the reason for the better computation overhead of $\textsc{FedAnil}$ is the leverage of lightweight computation techniques on the local enterprise and the server side. In addition, a theoretical convergence analysis for the $\textsc{FedAnil}$ model was presented. Convergence analysis showed that the model parameter obtained with $\textsc{FedAnil}$ converges to the optimal model parameter. The practical implications of this research can include the model's average communication overhead and vulnerability to data poisoning attacks (Label-flipping (Clean-Label or Dirty-Label)) and model poisoning (Sign-flip gradients or replacing gradients with a constant value). In future studies, an FDL-based model will be proposed to address the following challenges: Reducing communication overhead and addressing all three challenges of non-IID data (data type, feature, and label skew). The benefits of a communication-efficient FDL-based model can include increasing the model's scalability and efficiency and reducing the model's training time, complexity, and communication and computation overhead. Moreover, addressing all three non-IID data challenges (data type, feature, and label skew) can include reducing the convergence time and increasing model accuracy.

\printcredits

\bibliographystyle{model1-num-names}



\bio{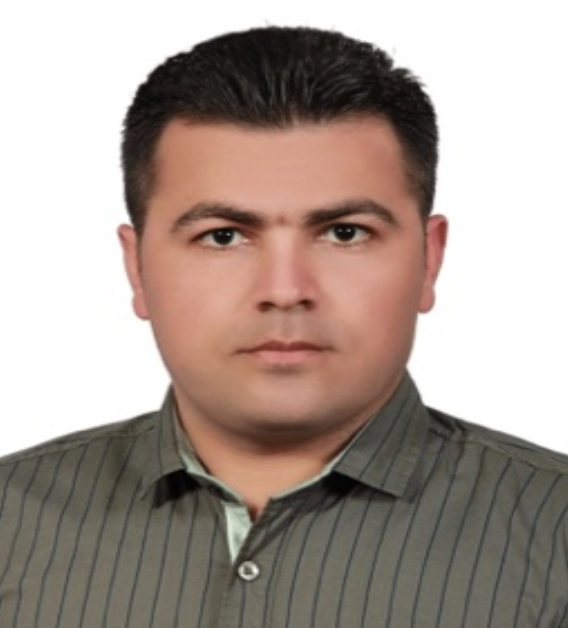}
\textbf{Reza Fotohi} is a Ph.D. Candidate in the Faculty of Computer Science and Engineering, Shahid Beheshti University. His research interests include Privacy-Preserving \emph{FL} (PPFL). Furthermore, he has published several papers in security domains in highly-ranked journals. His papers have over $1605$ citations with $27$ h-index and $33$ i10-index. Also, he is recognized as being among the World's Top $2$\% of Scientists (Stanford University Ranking, $2021$, $2022$ \& $2023$, \href {https://elsevier.digitalcommonsdata.com/datasets/btchxktzyw/3}{\color{blue}{Link}}).
\endbio

\bio{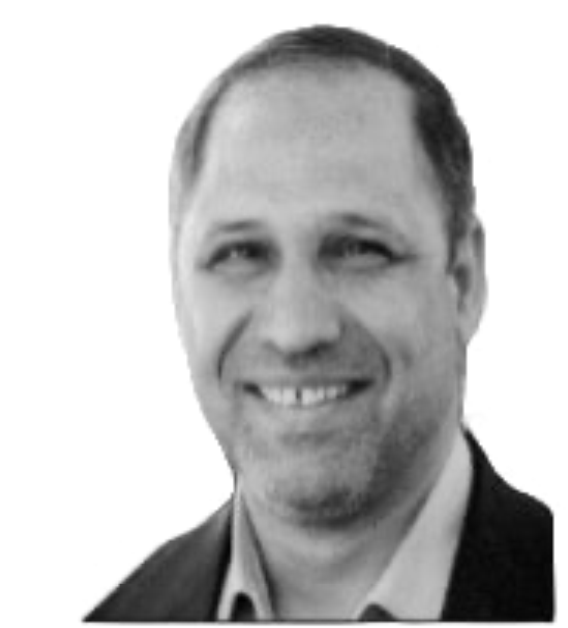}
\textbf{Fereidoon Shams Aliee} received his Ph.D. in Software Engineering from the Department of Computer Science, Manchester University, in 1996 and his M.S. from the Sharif University of Technology in 1990. His major interests are Software Architecture, Enterprise Architecture, Service Oriented Architecture, and Ultra-Large-Scale (ULS) Systems. He is currently a Professor at the Shahid Beheshti University. Also, Dr. Shams is heading two research groups, namely SOEA Lab, \href {https://soea.sbu.ac.ir/en}{\color{blue}{Link}} and ISA, \href {https://soea.sbu.ac.ir/en}{\color{blue}{Link}} at Shahid Beheshti University.
\endbio

\bio{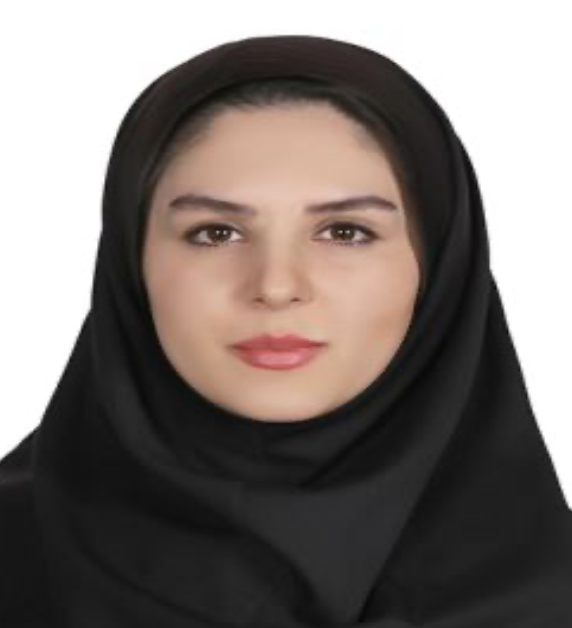}
\textbf{Bahar Farahani} received her Ph.D. and Postdoctoral degrees in Computer Engineering from the University of Tehran and Shahid Beheshti University, respectively. She is an assistant professor at Cyberspace Research Institute, Shahid Beheshti University. She authored several peer-reviewed Conference/Journal papers and book chapters on IoT, Big Data, and AI. Dr. Farahani has served as a Guest Editor of several journals, such as IEEE Internet of Things Journal (IEEE IoT-J), IEEE Transactions on Very Large-Scale Integration Systems (IEEE TVLSI), IEEE Transactions on Computer-Aided Design of Integrated Circuits and Systems (IEEE TCAD), Elsevier Future Generation Computer Systems (FGCS), Elsevier Microprocessors and Microsystems (MICPRO), Elsevier Journal of Network and Computer Applications (JNCA), and Elsevier Information Systems. Besides, she has also served on the Technical Program Committee (TPC) of many international conferences/workshops on AI/IoT/eHealth and is the Technical Chair of the IEEE COINS conference.
\endbio

\end{document}